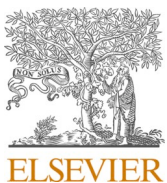
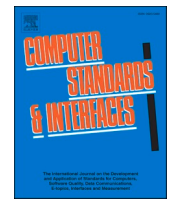

# Atmosphere: Context and situational-aware collaborative IoT architecture for edge-fog-cloud computing

Guadalupe Ortiz [a],[*], Meftah Zouai [b], Okba Kazar [b], Alfonso Garcia-de-Prado [a], Juan Boubeta-Puig [a]

[a] *UCASE Software Engineering Research Group, University of Cadiz, Puerto Real, Cadiz, Spain*
[b] *University of Mohamed Khider, Laboratory LINFI, Biskra, Algeria*

ABSTRACT

The Internet of Things (IoT) has grown significantly in popularity, accompanied by increased capacity and lower cost of communications, and overwhelming development of technologies. At the same time, big data and real-time data analysis have taken on great importance and have been accompanied by unprecedented interest in sharing data among citizens, public administrations and other organisms, giving rise to what is known as the Collaborative Internet of Things. This growth in data and infrastructure must be accompanied by a software architecture that allows its exploitation. Although there are various proposals focused on the exploitation of the IoT at edge, fog and/or cloud levels, it is not easy to find a software solution that exploits the three tiers together, taking maximum advantage not only of the analysis of contextual and situational data at each tier, but also of two-way communications between adjacent ones. In this paper, we propose an architecture that solves these deficiencies by proposing novel technologies which are appropriate for managing the resources of each tier: edge, fog and cloud. In addition, the fact that two-way communications along the three tiers of the architecture is allowed considerably enriches the contextual and situational information in each layer, and substantially assists decision making in real time. The paper illustrates the proposed software architecture through a case study of respiratory disease surveillance in hospitals. As a result, the proposed architecture permits efficient communications between the different tiers responding to the needs of these types of IoT scenarios.

## 1. Introduction

The Internet of Things (IoT) has become very popular in recent years; however, the term, introduced by Kevin Ashton in 1999 [1], is now more than two decades old. When it was first introduced the scope of the IoT was reduced to the supply chain, but today it refers to a wide spectrum of domains and is used in a broader sense, referring to the digital interconnection of objects and devices that can obtain and share information throughout platforms, providing added value to businesses, administrations and citizens.

The rise of the IoT has been accompanied by the improvement and reduction in costs of technology and communications, as well as an improvement in data analysis software; in particular, big data and streaming data processing, allowing for real-time decision making. Furthermore, it is not only a question of greater profits for companies; we live in a so-called fast-paced society and citizens want everything instantly and they want it to be adapted to their situation and specific context. Furthermore, they have already imposed a need to be part of smart cities, which requires the processing of large amounts of data and which are even more intelligent thanks to data democratization and citizen data sharing.

Therefore, the IoT requires not only the management of device data, but also the exchange of information between multidisciplinary platforms. The vast amount of data from multiple smart devices can and should be shared to provide added value and enable a broader global understanding of the domain in question. Collaborative IoT (C-IoT) takes account of IoT in the scope of heterogeneous domains and environments where sensors, gateways and services can interact at different levels; sensors provide their sensed data; gateways add intelligence and take action or communicate information at a higher level; lastly, services use the information provided by the gateway to improve the quality of life of people and business processes.

On the one hand, we are looking for low consumption protocols and software that allow a small device not only to send and receive messages, but also to have its own intelligence, thus going from a mere connected device to becoming a smart one. On the other hand, we want entire buildings and even entire cities to be connected, which requires higher bandwidth and massive, scalable data processing.

To accommodate the collaborative IoT, we will need to design software architectures that allow us to establish these communication channels, taking into account the capabilities and needs of each level of collaboration. We will have to design three levels of abstraction: (1) a lower level with the smart devices or edge devices that will obtain the information from the sensors and will be able to interact with the actuators, but that normally have limited computing resources; (2) a second level, in the fog, that will allow us to collect information from several edge devices and act as a gateway for communication with other gateways and with the third level; and (3) the latter being either the end user or being located in the cloud with its inherent high computing capacity.

Most of the approaches we find today focus on one issue or another; there are relevant approaches that center on ensuring the low resource consumption of the edge devices with their gateways [2,3] and others that see to the real-time processing of large amounts of data reaching the cloud [4]. However, few focus their efforts on providing a solution with three levels of granularity that provides the best benefits from all parties and particularly that permits appropriate processing in every tier and bidirectional communication between adjacent layers.

Hence, in order to address this issue, several Research Questions (RQ) arise; the first one is:

RQ 1. Can we integrate various specific software approaches that have proven to be efficient at the edge, fog and cloud, respectively, to provide a complete three-tier software architecture in which all three tiers can process information?

In recent years, we have seen several approaches that give good results at edge, fog and cloud levels. In particular, some of the authors of this paper have proposed the use of agent-oriented programming as an efficient solution for making intelligent decisions based on sensed data and data shared between the different edge nodes of the architecture [5]. Such agents will provide the edge nodes with intelligence, autonomy and capacity of cooperation and organization. The authors of the present paper have also made specific proposals for the processing of IoT data in context-aware cloud architectures [6,7] and in C-IoT environments [8] by combining fog with cloud nodes through the use of Event-Driven Service-Oriented Architectures (ED-SOA or SOA 2.0) and Complex Event programming (CEP). From our experience, we believe that the integration of the proposed solutions can provide a software architecture that covers the three levels efficiently. However, such previous proposals lack certain features that would significantly improve the usefulness of the proposed architecture. First of all, the agent-based software architecture [5] focused all the analysis for decision making in the agents' layer and the gateway acted as a sole intermediary and distributor of messages; thus, we lack data processing in the gateway that could improve decision making on the edge devices. Secondly, the C-IoT architecture involving cloud and fog nodes [8] lacks bidirectional communication between fog and cloud nodes, and therefore the cloud does not benefit from the knowledge and situations of interest detected in the fog. This leads us to the following research questions:

RQ 2. Would it be possible to have real-time data processing at every level of the integrated architecture as well as two-way communication between the edge and fog nodes and the fog and cloud ones by adjusting to system resources?

We certainly think so. By integrating the edge-specific architecture with the C-IoT-specific architecture we are providing real-time processing on all three levels of the architecture. In the case of the edge, the system has been tested efficiently with the software agents [5]. Regarding the cloud, resources are almost unlimited and we have seen that very high real-time processing ratios are achieved with the combination of SOA 2.0 and CEP [6]. Finally, as far as fog is concerned, we could see that processing with a SOA 2.0 combined with CEP was reasonably good [8]. Since the aim of this paper is to minimize the use of resources and improve the performance in the fog as well as providing the added value of coordinating communications with several nodes, we will study how to improve this performance because SOA may be an oversized solution for fog. On the other hand, in the initially analyzed architectures, communication between the different levels is carried out through message brokers, mainly through low consumption protocols such as Message Queuing Telemetry Transport (MQTT). Extending the use of the broker to provide bidirectional communication does not imply relevant impact on node performance.

RQ 3. If the answer to RQ2 is affirmative, would it be useful to have two-way communication among the three system layers and data processing in all of them in a broad range of IoT scenarios?

Again, we believe that the answer is affirmative: given the large amount of data obtained today from various sources (sensors, administrations, users, organizations, companies, etc.) and the proven benefit of real-time data analysis for decision making, we are convinced that many IoT scenarios will benefit from this architecture for the C-IoT. It will allow for the processing and exchanging of situations of interest detected at the three levels in order to have greater contextual and situational knowledge for decision making.

RQ4. Besides, if the answer to RQ2 is affirmative, can we ensure interoperability between layers and coordination between devices of the same or different layers despite the lack of agreement on communication standards for the IoT, and can we do it in a secure way?

Although there is currently no agreement on the standards to be used in terms of communications protocol and message payload format for the IoT, there is a trend towards the use of messaging brokers and low-power communication protocols such as Mosquito and MQTT [9], respectively. In terms of the message payload format, there is no established scheme or standard, although the use of the JSON format is very common. We believe that interoperability can be ensured by adding a transformation module, where necessary, to send the data to the corresponding node in the required format. Concerning coordination between devices of the same and different layers, on the one hand, we have to highlight that there are several widely used protocols for coordinating software agents; on the other, it is worth noting that message brokers also support coordination through their communication mechanisms; therefore, we expect a smooth coordination between devices. Finally, the security mechanisms used for communications are independent of the proposal being presented, and existing standard mechanisms for such communications can be applied, with particular attention to sensitive data.

As a result, the main contribution of this paper is Atmosphere, a software architecture that covers all levels of the C-IoT and meets the key requirements to address decision making from contextual and situational heterogeneous data in an efficient way.

The proposal will benefit, on the one hand, from the intelligence, autonomy and low consumption of software agents in the edge tier; on the other hand, having a gateway in the fog will allow us to centralize the information from various devices related to each other and process their relevant data and those obtained from the cloud to enhance decision making. Finally, the cloud node will feedback and be fed back by the gateway with further situations of interest detected when processing data from several sources. Despite the lack of standards, the use of protocols and data formats common to IoT information exchange in this proposal will facilitate interoperability between the layers and components of the proposed solution. Besides, we have illustrated the proposed software architecture through a scenario for hospital patients with respiratory diseases, where third parties such as pharmaceutical or research laboratories interact with the system.

The rest of the paper is organized as follows. Section 2 provides background information on the technologies and paradigms involved in this paper. Then, Section 3 provides two specific examples of C-IoT scenarios. Afterwards, Section 4 describes the proposed Atmosphere architecture; it is initially described at a higher level for better understanding, then communication between the different tiers and the software architecture in each tier are explained. Section 5 describes how the architecture would be deployed in a case study based on one of the illustrative scenarios in Section 3, providing some samples of agent rules and event patterns to facilitate the comprehension of the system. Section 6 shows the results of the architecture performance evaluation. Section 7 discusses and compares existing related work to our provided proposal and Section 8 brings further discussion on the proposal itself. Finally, conclusions and future work are presented in Section 9.

## 2. Background

This section firstly introduces the relationship between edge, fog and cloud computing and, secondly, it places it within the context of the C-IoT. Then, Ed-SOA is introduced, followed by an explanation of CEP and message brokers. Afterwards, context and situational awareness are explained. Finally, agent-oriented software is defined.

### 2.1. Edge, fog and cloud computing

It is remarkable that we can find various definitions for edge and fog computing in the literature and that, among these definitions and descriptions, there are sometimes overlaps and unclear boundaries.

Fog computing and edge computing seem similar, as both involve bringing intelligence and data processing closer to the data source. However, in fog computing, intelligent processing is expected to take place in a gateway in the local area network where the endpoint nodes submit data to and from which they receive processed data back, whereas, in edge computing, intelligent processing is expected to take place in final devices near the edge of the network [10].

When we address edge computing we are referring to a hardware infrastructure with computing capacities, which is generally located near the devices making use of said infrastructure. In the IoT scope, normally so-called IoT devices provide such computing capacities and are named edge nodes, edge devices or smart devices, even though, depending on the granularity of the particular scenario, the devices may have higher or lower computation capacity. Having such computing capacities in the device itself provides several advantages: the first one is that we can preprocess the data obtained in the edge device and communicate such information to external devices only when relevant or necessary, saving edge device resources and improving latency; the second one is that the device can make its own decisions, if so programmed, particularly useful when there is no network connection.

With millions of interconnected devices, the generated data can be significantly extensive. Therefore, it is not possible to use the traditional hardware environments and software tools to manage and process the data with an acceptable response time, but fog computing has the potential to provide better performance for real-time services [11]. In this sense, fog computing responds to the need of an intermediate layer between the cloud and the smart end devices, where the fog provides support for low latency real-time analytics of data that cannot be processed in the end-device as well as to interrelate and process together data from different devices and to provide support to communications with external services. Edge devices are usually limited in computing and storage but may process certain data the instant they sense them and may make quick decisions that do not depend on other devices' data [12, chapter 3]. Therefore, the use of fog computing does not involve either doing all the analysis on the edge or not doing analysis on the edge, but promotes doing as much processing on the edge as is possible and necessary within realistic constraints and processing the rest of the information in the fog. This way, some data are processed by applications running on edge devices within the network and the rest of the data is processed by a node in the fog, which is also responsible for intelligently orchestrating all connected devices, as well as, if necessary, devices in the cloud [13].

Indeed, one of the main objectives of the combined use of edge and fog computing is to have intelligent sensing: submitting only useful data based on the knowledge and intelligence available locally at the edge device to be processed at the fog [14].

Thus, fog computing provides certain advantages such as fewer resource requirements for edge devices, improved context awareness, low latency real-time processing and lower bandwidth requirements thanks to only having to submit the necessary collected data to the cloud instead of all of it, as well as facilitating collaboration between several edge devices [14].

Why do we still need a cloud? Because normally these fog devices will also have limited resources, and they are expected to be close to the final devices, where there most probably will be space restrictions. Besides, the information from many fog devices might be necessary to make higher level decisions, therefore they have to be submitted to a central computing node (in the cloud).

Please note that sometimes the edge node can act as a fog node (with enough computing capacities and submitting relevant information directly to the cloud); other times, the suitable architecture might include both edge and fog nodes, as well as, of course, cloud ones.

It is also worth clarifying that sometimes the terms edge and fog are used indistinctly. The reason is that at times having a fog or not will depend, as previously explained, on the particular scenario; we might have an edge device with enough capacity to act as a fog node and directly communicate with the cloud. This is why for some authors, edge and fog devices might be considered as the same thing [15], since they are opposite to cloud computing nodes in the network connection [16]. However, in this paper we distinguish three tiers —edge, fog and cloud—, as previously motivated [10–14].

### 2.2. Collaborative Internet of Things

C-IoT [17] considers the IoT in the scope of heterogeneous elements and domains, where sensors, gateways and services potentially interoperate at different levels; sensors provide their sensed data; gateways add intelligence to them and undertake actions or report information to a higher level, and services utilize the information provided by the gateway to enhance people's life quality or business processes.

Collaboration on the IoT takes place on three concentric levels: the innermost level would be the individual, the intermediate level would be the industrial level and finally the outer level would be the infrastructure. At all these levels, we can find a pyramid of applications that allow the materialization of the C-IoT: at the base of the pyramid we would have everything related to the sensor, at the intermediate level, the gateway and, finally, at the top of the pyramid, the services offered to the end user [17].

For example, in the application domain of a pandemic, sensors of medical devices, patient data, medicines, etc., would be placed at the base of the pyramid; in the gateway we could have patient histories, research results and known effects of the treatments; thirdly, in the cloud there would be a large bank of services to access treatments by age group, results, patient evolution depending on the treatment, affected by geographical area or admitted to each hospital, etc. We therefore have individual data (patients), industry data (treatments) and infrastructure data (hospital center information).

Having said that, we can easily identify the previously explained terms, edge, fog and cloud with the three levels in the pyramid: sensing, gateway and services, respectively.

Besides, what will give us great added value will be the combination of data obtained from sensors or information systems with data provided by citizens, and will allow us greater control and monitoring of various C-IoT scenarios, such as e-health [4], smart cities [18], Industry 4.0 [19], etc. To do this, we will need to be able to process heterogeneous data from various sources in real time. In order to be able to process heterogenous data coming from several sources we will require a software architecture that permits it; this leads us to introduce Event-Driven Service-Oriented Architectures in Section 2.3. Furthermore, to be able to perform data analysis streaming in real time, we will also require additional technologies, which brings us to explain complex event processing and message brokers also in Section 2.3. It is worth pointing out that when talking about real time throughout this paper, we refer to quasi real time. This term differs from the strict traditional definition of real-time computation, where real-time responses are expected to be received in the order of milliseconds or even microseconds. Generally, the term quasi real time refers to a short-time response from a system according to its needs, it might be in the order of milliseconds or maybe in seconds. For instance, as we will later see in the case study, if we need to warn a doctor to assist a patient a millisecond difference in the response time is not noticeable (some minutes delay would). Therefore, such systems respond rapidly to the occurring events but do not require strict under millisecond response.

Note that the gateway in such smart systems for C-IoT has to provide a number of relevant features [17]:

1. It has to be context- and situational-aware to allow service customization.
2. It needs to be able to make decisions on its own, without the need for sending the data to the cloud, which would delay the response.
3. Given a certain context, it must be able to anticipate certain actions, since it knows the situation and the user from a pre-existing history.
4. It must be autonomous, be able to interoperate and collaborate with other nodes in the system and make decisions based on a series of rules of reasoning facilitating communication between different nodes.

According to feature number 1, we will need to introduce context and situational awareness —see Section 2.4—. In order to fulfill features 2 and 3, we will see how CEP, isolated and combined with a SOA 2.0, together with a message broker could be used —see Section 2.3—. Finally, autonomy —feature 4— will be provided by an additional technology, an agent-oriented software, introduced in Section 2.5.

*2.3. Event-driven service-oriented architecture and complex event processing*

SOA represents a paradigm for the design and implementation of loosely coupled distributed systems where services are the main implementation mechanism. These architectures deliver easy interoperability among third-party systems in a flexible and loosely coupled way. This way, costs are reduced when system modifications or improvements are required, since the system will be easier to maintain and evolve [20].

Success and growth in the use of service-based systems require a new service infrastructure that allows us to interconnect and maintain applications flexibly. This infrastructure must support well-known web service standards and provide support for a message middleware [20]. These requirements are fulfilled by an Enterprise Service Bus (ESB). An ESB facilitates interoperability between various applications and components through standard service-oriented interfaces, facilitating the development of complex systems based on a SOA and allowing applications to be offered as services in the ESB. The bus provides other additional advantages, among which we should highlight the assurance of system scalability.

ED-SOA, or SOA 2.0., evolves from traditional SOA. Unlike traditional SOAs where communications are mainly carried out through remote procedure calls, in SOA 2.0. communication between users, applications and services is mainly conducted by means of events [21]. This new paradigm also benefits from the use of an ESB. In this case, the ESB will allow us to integrate various heterogeneous data sources and invocations to various distributed systems and applications [22].

Despite all the advantages provided by SOA 2.0, this type of architecture might not be ideal to analyze and correlate large amounts of data in terms of events in real time. To meet this requirement, it is necessary to integrate CEP [21], a technology that allows the capturing, analyzing and correlating of a large amount of heterogeneous data (simple events) with the aim of detecting relevant situations in a particular domain [23]. This situation of interest detected by an event pattern is named *complex event*. The software that permits the analysis of streaming data according to the defined patterns in real time is the CEP engine. Among the complex event engines, Esper stands out due to its maturity and performance, as well as the wide coverage of its EPL event pattern definition language [24].

In addition, when large amounts of events are received, and have to be processed by the system, it is convenient to integrate a new element that facilitates such communications and the management of the input data. To do this, it is convenient to use a message broker. Message brokers implement an asynchronous mechanism which allows source and target messages to be completely decoupled, as well as allowing the storage of messages in the broker until they can be processed by the target element.

There are brokers and protocols specifically designed for devices with limited resources. In particular, the MQTT [25] protocol was proposed as a light protocol implementing a publish/subscribe mechanism for Machine-to-Machine (M2M) communication; the broker Mosquitto [26] implements MQTT and is being frequently used in the scope of IoT nowadays.

*2.4. Context and situational awareness*

Dey et al.'s context definition in [27] is specially well-known, where "*Context is any information that can be used to characterize the situation of an entity*"; such an entity can be almost anything —an object, a device, a citizen, et cetera—, which can be useful to improve the interaction between the end user and the computer system or application, as well as to enhance the functionality of the application itself. Context will be specific to each application and system domain and even to each system use.

*Context awareness* supports the fact that context information, obtained from the system environment, is properly used by the system to improve its quality in terms of user satisfaction, and accuracy in terms of the functionality of the system itself. That is, a context-aware system is expected to use information such as location, company, known social attributes, personal status, domain-specific known facts and other available information to anticipate system needs so that more customized systems can be delivered. Therefore, a system is context-aware if it uses the context to provide relevant customized information and/or services to the user or to the system itself, adapting its system behavior to the particular needs of the specific user or system [28]. Context and context awareness have become a key issue for decision making in general and for real-time decision making in particular [29].

This leads us to the concept of *situational or situation awareness*, originally defined by Endsley [30] as "*the perception of elements in the environment within a volume of time and space, the comprehension of their meaning, and the projection of their status in the near future*". If we extend such a concept to computing situational awareness, we can think of it as the understanding of the events related to the entity and its context in a given situation. As a result, the situation would be the result of context information pieces being interpreted and aggregated, in order to obtain more meaningful situations of interest.

*2.5. Agent-oriented software*

In 1993, Shoham defined agents as "*an entity whose state is viewed as consisting of mental components such as beliefs, capabilities, choices, and commitments*" [31], where the choices or actions will be based on the agent's decisions, the latter being determined by the agent's beliefs according to its and other agents' capabilities.

More recently, in 2003, Silva et al. defined the concept of agent and multi-agent systems (MAS) in the field of software engineering and distributed systems [32], where agents are expected to represent autonomous components in software systems. An agent's behavior is defined based on its plans and actions, its characteristics and the interaction with other agents. In addition, they are considered proactive thanks to their autonomy to perform tasks. Besides, they are adaptive, since they adapt their status and behavior by responding to messages from other agents.

In MAS, the agents are autonomous and their characteristics allow them to interact with the environment and other agents. A MAS is considered to be an object-oriented system that is associated with an intelligent meta-system. In this manner, an agent is seen as an object that has an intelligence layer, which comprises a series of capabilities such as uniform communication protocol, perception, reaction and deliberation, all of which are not inherent to objects. In MAS environment, an Agent Communication Language (ACL) is required to send a message to any agent, FIPA ACL being a well-known one [33]. We will also require an agent platform and a well-known platform is JADE [34], the Java Agent Development Framework which permits developing agent-oriented applications according to the FIPA patterns. JADE is based on the coexistence of several Java Virtual Machines (JVMs) distributed over multiple computers: each computer will have a JVM, with a container of agents; the JADE platform runs at the main container.

*2.6. State of the Art for Edge-Fog-Cloud Software Architectures*

Although we can find multiple works that deal with edge computing, fog computing, edge-fog, fog-edge and so on, the vast majority of them approach it from a network communications and communication protocols perspective. We have found few works that study the software architecture that allows developers to take better advantage of the data flowing between the three layers —edge, fog and cloud—. In this section, we will summarize some of their shortcomings that motivate our proposal; a more exhaustive analysis of such and other proposals can be found in the related work section.

Among the proposals found that feature a software architecture with the three layers mentioned above [2–4,35–37], we have noticed that most of them provide computing mechanisms for both the fog and the cloud layer, but rarely provide computational mechanisms for the edge, and are limited to sending the data obtained by the device at that layer. Thus, these proposals [2,35–37] do not take advantage of edge computing being able to facilitate faster decision making directly between edge devices when the scenario demands it.

Another shortcoming of many of these proposals is the fact that communication between layers is not always in real time [2–4,35,36], which limits the ability to make timely decisions in the specific application domain. Likewise, communication between edge nodes and end users is rarely facilitated —only in [2,36]— which would allow faster communication for highly relevant situations detected at the edge that may require an immediate response from the end user.

Finally, we have already pointed out in the introduction the benefits that can be obtained from real-time bidirectional communications between the three layers; however, we see that only one of the mentioned proposals implements bidirectional communications [37], which, however, is one that does not provide edge computing.

For all these reasons, we believe it is necessary to propose a new software architecture that provides all these added benefits to facilitate faster and more effective decision-making. In order to choose the technologies that will form part of each level of the architecture, we have drawn on our previous experience, through which we have observed the usefulness and efficiency of using agents at the edge and CEP in the fog and CEP-enhanced SOAs in the cloud. This is not only based on our own experience, we see that several proposals are emerging that propose the use of agents as an efficient solution for decision making at the edge in the scope of IoT [38–41], as well as those that include agents in the fog [42,43] and of course proposals that take advantage of the great efficiency of CEP for real-time data analysis in the cloud [44,45]. However, we believe that they do not address the joint use of all these technologies in a multitier edge-fog-cloud software architecture, which is what we contribute in this paper, taking a step further in the development of multitier software architectures for the IoT.

## 3. Illustrating scenarios for C-IoT

In this section we provide two illustrative examples of C-IoT scenarios; the first one is focused on hotel amenities and energy consumption monitoring and the other on hospital patient monitoring.

*3.1. C-IoT scenario for hotel amenities and energy consumption monitoring*

Let us imagine a large hotel totally automated with new technologies. We could have several IoT devices in each room; let's suppose we have the following devices:

- Air conditioning/heating: this IoT device would detect the temperature in the room and activate air conditioning/heating to adapt to the user's comfort temperature. The user can program its operation, for instance, according to the time, the number of people in the room and the activity (for example, if he/she is taking a bath).
- Bath/Jacuzzi: this device can also be programmed to the user's favorite temperature; it could also be programmed at a certain time and its information could be valuable not only for the air conditioning/heating device, but also, for example, for the background music device.
- Background music: when the bath is running, the background music device is switched on and relaxing music is played. Or perhaps, depending on the time of day, if the bath is used in the morning it could play upbeat music, and relaxing music in the evening.

We have some observations to make regarding communications. On the one hand, we see that several edge devices will require information obtained from other edge devices in the room: for instance, the air conditioning and background music devices should know when the bath is activated/deactivated. On the other hand, communication with higher level actors is also required. First of all, the user must have a way to program all his/her tastes through an interface, which would be translated into conditions for the edge devices. But the hotel manager might directly program a whole hotel floor according to the tastes of a regular customer or a group of regular customers. The hotel manager can also receive information from all the rooms, for example, on each room's energy consumption levels or temperature, which could result in rules for energy saving or fire detection.

*3.2. C-IoT scenario for hospital patient monitoring*

Let us imagine another scenario, the hospitals in a certain region. Suppose we find ourselves in the context of a time of year with a high incidence of contagious respiratory diseases. In this scenario, the hospital could be equipped with a set of devices, such as:

- Access devices: to enter the rooms of patients with such respiratory diseases, you can only access with ID cards that, in addition to denying passage to unauthorized visitors, keep information about which people and when they have accessed the room.
- Ventilators: ventilators are equipped with a series of sensors that detect the $O_2$ and $CO_2$ blood levels as well as the patient's flow and proximal pressure.
- Windows: windows are equipped with actuators to open and close for room ventilation, when necessary.

- Exterior emergency light: rooms are equipped with an exterior emergency light to indicate that a doctor needs to come to the room urgently.
- Interior emergency light: rooms are equipped with an internal emergency light to indicate that attention is needed due to an event that is happening outside the room.
- Control panel with patient local history: doctors would insert information about administered medication and any reactions to the medication into it. Other doctors can also see the aforementioned information in it, as well as relevant messages related to the patient history.

Concerning communication between edge nodes, if the $O_2$ blood level detected by the ventilator is below a certain threshold, the external light of the room would be immediately activated, the door would be unlocked and the room would be ventilated. If the number of rooms with the external light on is over a set threshold, the internal lights of the remaining rooms are illuminated for a short period of time so the doctors inside them are aware of this fact.

In the hospital, we could have a control center that receives the relevant information from all the edge nodes in each room on every floor. If the number of rooms on the same floor with an event that has activated the outside light happened to go above a certain threshold, the control center would receive an alert and immediately send more doctors to that floor.

We could also have access to the patient history from the control center. The control panel in the room would send all recorded information about medicines and reactions to the aforementioned center to be processed, thanks to which we could detect if the patient is being given some medicine that previously gave him/her a reaction and send a message back to the control panel at the edge. We could also detect if the same medicine is causing other patients any side effects, information that could be raised to the cloud, which research laboratories might be subscribed to. In a scenario such as the current one, with a pandemic and a massive shipment of vaccines to different parts of the world, laboratories would benefit from knowing about possible side effects of a certain severity in real time, not only because the knowledge of side effects in general could lead to measures. This would be particularly useful because receiving these alerts in real time could lead to the same adverse effect arriving from different hospitals on the same date, raising suspicions about a defective batch and leading to measures being promptly taken, such as stopping the production of the medicine and urgently alerting other recipients who have received vaccines from the same batch, etc.

As we have seen, there are several scenarios where an architecture combining edge, fog and cloud computing might be necessary. In each scenario, we might require some adjustments in the architecture with regards to the number of edge nodes, fog gateways and cloud nodes. In the following section, we will explain the flexible architecture suggested for such scenarios.

## 4. Atmosphere architecture

Atmosphere is a collaborative architecture for the IoT, composed of several edge, fog and cloud nodes which collaborate in a service and agent-oriented ecosystem through the cooperation of software agents and CEP.

We will have three layers in the architecture, such layers will conceptually focus on sensing data, acting as a gateway and offering services and will be materialized in the corresponding edge, fog and cloud nodes. Fig. 1 shows an illustrating architecture where one cloud node, two fog nodes and five edge nodes have been included. However, the architecture could have additional nodes of all types (to illustrate this, some extra fog and edge nodes are also represented in smaller sizes). In any case, the architecture is expected to have more edge nodes than fog ones and in turn more fog nodes than cloud ones.

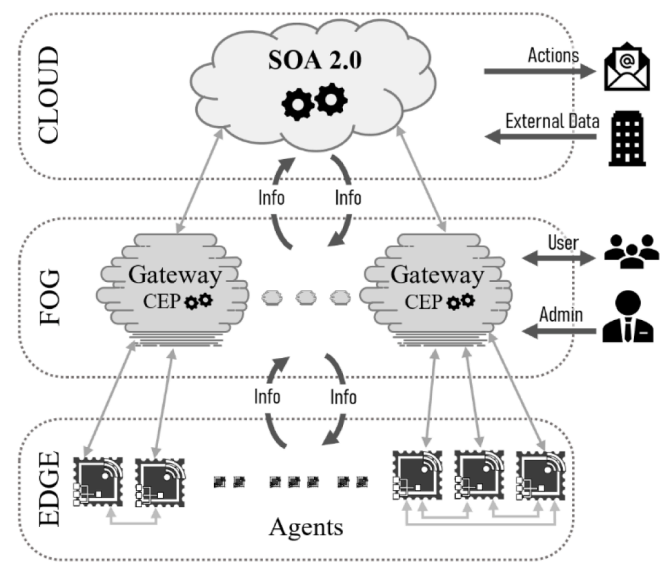

**Fig. 1.** Atmosphere architecture high level description.

### 4.1. High level layer and component description

As previously mentioned, our architecture will have edge nodes, fog nodes and cloud nodes, as explained in the following paragraphs.

**Edge nodes**: we must remember that IoT devices, represented as edge nodes, tend to have limited computing resources, that is why (1) the software they incorporate must consume few resources in execution and (2) communications through the Internet must be limited as much as possible to save resources. With this objective, we have deployed a software agent in every edge node, to filter the information that does need to be transmitted from the one that will only be used internally, since agents' consumption is low and they have autonomy and enough intelligence and decision capacity. The agent's intelligence is programmed based on a series of behavioral rules [46]; this way, the device behavior will be able to adapt to the environment, making decisions according to the information obtained from its sensors and to the information obtained from other edge nodes connected to the network and from the fog node. The agents will be able to communicate with each other, processing and analyzing the information sensed and received from other actors and making the corresponding decisions without the need to contact the user.

**Fog nodes**: these are devices that will act as gateways between the IoT devices on the edge and (a) the user and (b) the nodes in the cloud [47]. On the one hand, all communications of the software agents with the end user will take place through a message broker in the gateway: edge devices can send messages to a topic in the broker which the user is subscribed to. On the other, interactions between the fog and the cloud will also be done through other message broker in a bidirectional way. Said broker will be connected to a CEP engine in the fog node to process the received data and detect relevant information for the other party. That is, (1) the cloud sends information to the fog, the CEP engine processes such information and if a pattern is detected, then it submits the relevant information to the edge; (2) the edge sends information to the fog, the CEP engine processes the said information and if a pattern is detected then it submits the relevant information to the cloud. Of course, a pattern can be detected according to the information received both from the edge and the cloud.

**Cloud Node**: The cloud node has a two-way communication system with nodes in the fog. On the one hand, the cloud node subscribes to the message topics of the nodes in the fog, so it can obtain information from all these nodes and make higher level decisions. This information is acquired along with other heterogeneous data sources and processed through a CEP-enabled SOA 2.0. The CEP engine might detect situations of interest either for third parties or for fog nodes, and submit them to the corresponding message brokers which such actors might subscribe to. It is important to remember that there could be more nodes in the cloud that might be communicating with other nodes in the cloud at the same time, but for simplicity's sake we are going to represent only one.

To set up the system, first of all, the edge devices will be programmed with the software agents and the corresponding rules with the actions they may perform depending on the information received from the sensors, the other agents, the gateway and the user. These rules will not only include actions such as turning on the air conditioner, but also which information might be broadcasted to the remaining agents as well as which information should be submitted to the user or gateway topic to be processed in the CEP engine. Secondly, we will deploy and configure the message broker and the CEP engine in the fog node, which will receive data from the edge, other fog and cloud nodes, as previously explained. For this purpose, the actors will be subscribed to the corresponding topics. Besides, the patterns designed in the CEP engine will be deployed. Finally, we will also deploy a message broker and SOA 2.0, with an integrated CEP engine, in the cloud. We will also subscribe the message broker to the data sources of interest, as well as to the fog nodes corresponding to the scenario in question, and the designed patterns will be deployed. We will also deploy other desired actions for when a pattern is met (notification submission, service offering, etc.). Examples of such rules and patterns will be given in Section 5.2.

If we focus on the first illustrative scenario, we can visualize the devices in each room as edge nodes. On the other hand, we can visualize a fog node on each hotel floor, and finally a node in the cloud for the hotel itself, as represented in Fig. 2. What is grouped by each node in the fog or in the cloud obviously depends on the particular case study and the granularity we might want to give each layer. We could have a small hotel with only one node in the fog. And perhaps, we might have a hotel chain owner, with a fog node for each hotel and a cloud node for all his/her hotels. Sometimes, granularity does not depend on the size, but it can depend on the functionality that we want to give each layer, for example.

Regarding the second illustrative scenario, we can expect the devices in each room to be edge nodes. On the other hand, we can visualize a fog node at the hospital control center or maybe on each floor, depending on the size of the hospital. In Fig. 3 we represent the architecture with two fog nodes, one communicating with the edge nodes from all the rooms and the other for patient admission and staff monitoring. Finally, a node in the cloud could be from the Ministry of Health or from a medical society. Such cloud nodes could collect information from many sources, such as pharmaceutical laboratories, pharmacies, and so on. For instance, it could be detected thanks to the information provided by pharmaceutical laboratories and pharmacies whose demand for a particular medicine, which is administered at the hospital, is growing out of control, and there could be a lack of stock in a short period of time. That information should reach the hospital fog node and be transmitted to the doctors by the hospital control center, as well as being shown in the edge control panels of patients with respiratory diseases.

Again, granularity and the number of nodes will depend on many factors: hospital size, the number of hospitals supervised by a general hospital and so on.

### 4.2. Communications and insight on the components in atmosphere

In this section, we will give further details on the internal architecture of each layer in Atmosphere and explain all the possible communications that can take place between its components. Remember that the edge will not communicate directly to the cloud, but will do so through the fog node. All these communications are made using a lightweight messaging broker through the MQTT protocol.

**Edge nodes insight and communication**

In the edge nodes, we are going to have the following modules for every agent:

- Environment components. They consist of sensors and actuators of the domain in question. These objects are able to communicate to provide data (measurements) or perform commands.
- Rules Database (DB) component. It collects the agent's knowledge and contains rules that help the agent to make decisions.
- Analysis module. It checks if the rules are satisfied based on data obtained from sensors and data sent by other agents. These rules are defined in the rule database. If the agent detects that a rule is satisfied it sends information to the parties involved (other agents, and/or the fog node).

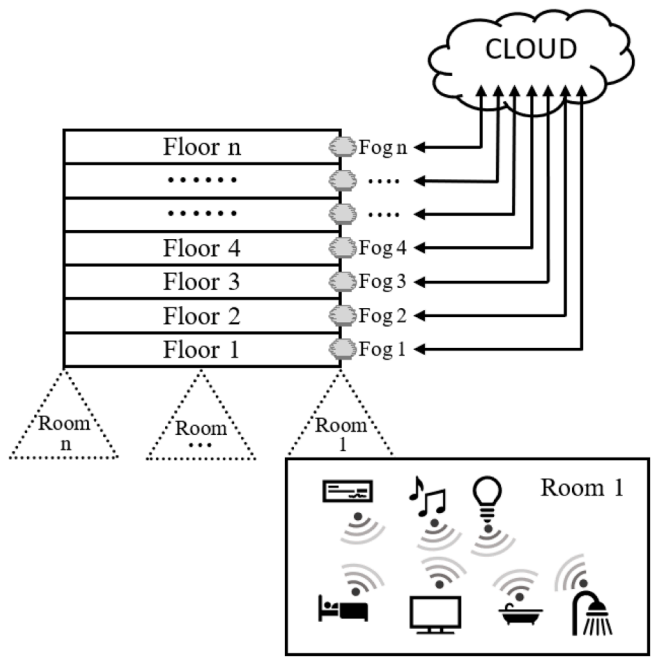

**Fig. 2.** Representing atmosphere architecture in a hotel amenities and energy consumption monitoring scenario.

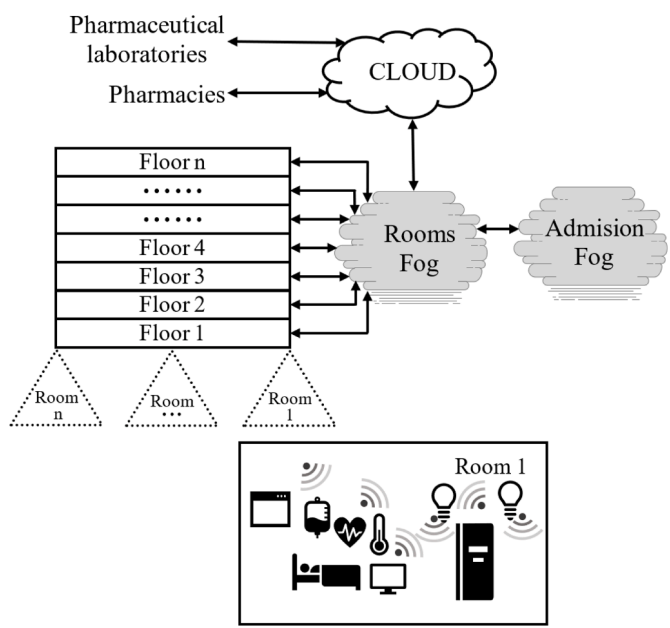

**Fig. 3.** Representing atmosphere architecture in a hospital patient monitoring scenario.

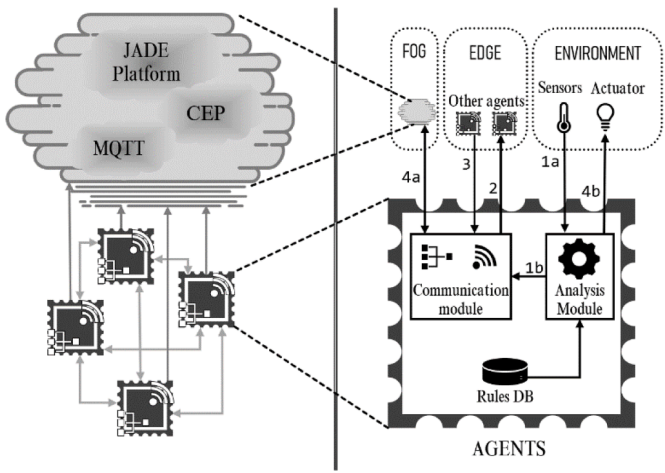

**Fig. 4.** Insight on edge nodes and their communication with the Fog.

- Communication module. It provides all the interaction mechanisms of the agent with the other agents and the fog. The agents communicate with each other and with the JADE gateway in the fog through the standard FIPA ACL communication language.

Therefore, each edge node will be driven by an agent. The agent will access the information sensed from a specific sensor and/or will activate/deactivate the actuator in question. It will have a set of rules to detect information of interest to be submitted to other agents and/or to the fog node. Besides, it will communicate with the latter thanks to the communication module. All communications are supported by the JADE platform, which we have used to implement our agent-oriented software.

Therefore, there are several communications that an edge device might be involved in, as shown in Fig. 4

1. First of all, (1a) the sensors in the device will sense the information from the environment and (1b) the analysis module will submit the sensed information to the communication module in the edge device.
2. Secondly, when necessary, according to the rules programmed in the agents, the sensed information will be transmitted to the remaining interested edge devices within the same network.
3. Analogously, the device will receive information from other devices in the network.
4. Besides, depending on the rules programmed in the agent and the decisions made, the edge device (4a) might submit information to the message broker in the gateway and/or (4b) activate an actuator.

**Fog Nodes Insight and Communications**

In the fog nodes we are going to have the following components:

- Component for the management and coordination of edge agents: JADE platform. It offers flexible and efficient messaging, managing private incoming message queues for each agent. Agents can access their queue in a variety of ways. In addition, communications and coordination between them is done through message passing following the ACL standard.
- Real-time data analysis component: Esper CEP engine. It is a specific software capable of real-time analysis of the data received in the form of events. The patterns to detect relevant situations of interest in the domain in question should be deployed in advance in this engine —some examples of CEP patterns can be found in Section 5.2—. In the case of the fog node, the CEP engine (1) will receive the data submitted from the edge, the cloud and/or other fog nodes, (2) will check if any of the predefined and deployed patterns is triggered by such data and (3) will submit the detected relevant situations either to the edge, the cloud or other fog nodes depending on the particular pattern being triggered —further details are given below—. Please note that messages received in the CEP from different sources may include the same or different fields to be processed. In case they include the same fields, an additional field may be included to identify the source in order to better coordinate their processing.
- Component for messaging incoming and outgoing data to the node: Mosquitto message broker. A message broker is software that allows applications and systems to communicate with each other and exchange information. The message broker can enable the translation of messages between different messaging protocols, allowing third parties to communicate with each other, regardless of their languages and implementation platforms. In our fog node, as we will explain below, it will permit bidirectional communication between the agents in the edge and the CEP engine in the fog, between the latter and the SOA 2.0 in the cloud and between CEP engines of two fog nodes.

Therefore, on the one hand, the JADE platform that supports the agents distributed among the edge nodes, and on the other hand, a CEP engine. Since all messages received in the fog are formatted inside our own architecture, they will already be in a homogeneous format, thus we will avoid the need for a SOA 2.0 and an ESB and will only require the CEP engine. In the event that our own sources could not be programmed to send data in a homogeneous format, we could include a format adapter, specific to our devices, prior to sending it to the CEP engine. We will also have a message broker to offer input and output message topics to the fog node. The relevant events detected by the edge nodes will be sent to the entry topic through the JADE coordination mechanism in the gateway; in that same topic, events of interest sent from the cloud or from other fog nodes will be received. All data received in that topic will be processed by the CEP engine which will upload new situations of interest detected to the cloud, other fog and edge nodes, as programmed in the pattern through a new output topic. In addition, the user can send new conditions to the edge nodes through a topic and receive information of interest by the same means. Therefore, communications of the fog node, the gateway, with the other participants would be as represented in Fig. 5:

Thus, the behavior is as follows:

1. The fog node will receive the following information:
   a. The fog node will receive relevant information submitted from the agents at the edge through a message topic.

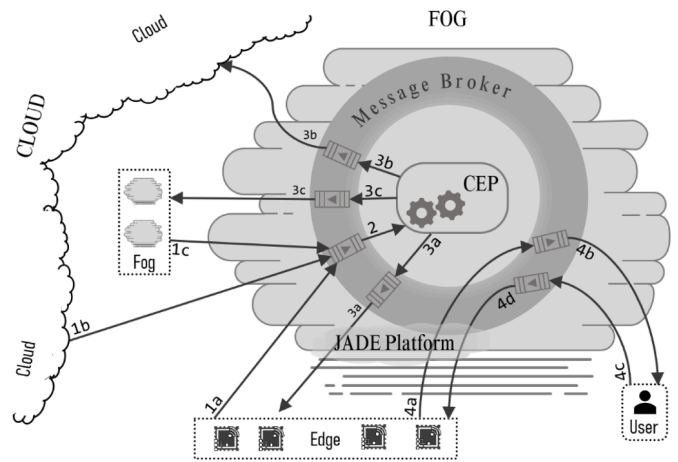

**Fig. 5.** Insight on Fog Nodes and their communication with the edge and the Cloud.

   b. Also, the fog node will receive relevant information to be processed from the cloud. This information will be received through the same message topic.

   c. Finally, the fog node might obtain relevant information to be processed from other fog nodes, which will also be received through the message topic.

2. As the information reaches the broker, it is sent to the CEP engine.
3. As the information is processed in real time in the CEP engine, various situations of interest can be detected. These situations of interest can be sent to three output topics, according to the action associated to the detected pattern, and may be relevant to the edge, the cloud and/or other fog nodes. Thus, the fog node will send the following output information:

   a. Relevant situations of interest to the edge nodes.

   b. Relevant situations of interest to the cloud nodes.

   c. Relevant situations of interest to the other fog nodes.

4. Finally, there will be direct communication with the user, which does not go through the CEP engine:

   a. The agents at the edge will submit relevant information to the user, according to the rules programmed in them; such information is submitted through a message topic.

   b. The user will be subscribed to that topic and therefore receive such information through it.

   c. Similarly, the user will submit new conditions, rules or actions to the edge through the message topic.

   d. Finally, the edge nodes will receive such information through their subscription to the topic.

**Cloud Node Insights and Communications**

In the cloud nodes we are going to have the following components:

- Component for the pre-processing of the information and submission to the CEP engine: Mule ESB. An ESB is in charge of routing and facilitating communications in a SOA. In the proposed architecture, the bus channels the following communications: reception of simple events from the message broker, event transformation and routing to the CEP engine and to a database if necessary, complex event reception of the complex events detected by the CEP engine, complex event routing to the relevant message broker.
- Real-time data analysis component: Esper CEP engine. As previously mentioned, it is in charge of analyzing the data received in form of events in real-time. The patterns to detect relevant situation of interest in the domain in question should be deployed in advance in this engine. In the case of the cloud node, the CEP engine, first of all, will receive the data submitted from one or several fog nodes and external systems relevant for the domain in question. Then, it will check if any of the predefined and deployed patterns is triggered by such data. Finally, it will submit the detected relevant situations either to the fog nodes through the message broker or with other protocols and/or formats to other interested parties out of the system —for instance, through SMTP, Firebase notification, HTTP service requests, etc.
- Component for messaging incoming and outgoing data to the node: Mosquitto message broker. In our cloud node, as we will explain below, it will permit bidirectional communication between the CEP engine in the fog and the SOA 2.0 in the cloud, as well as permitting the input of events from external systems and output of notifications to the latter.

Therefore, the cloud node is implemented as a SOA 2.0, which receives information from one or several message topics. The ESB in the SOA 2.0 is in charge of homogenizing the heterogeneous information received from all data sources, some of them external to the system (1) and others coming from fog nodes within the system (2). Once it is well formatted, it is submitted to the CEP engine where relevant situations of interest are detected. Such enhanced information is submitted to a message topic which the fog nodes in the architecture and/or external third parties can subscribe to. Besides, the fact that a SOA 2.0 is used allows us to introduce additional actions when a situation of interest is detected, not only sending the information to a message topic (3a, 3d), but also offering additional services (3b) and sending alerts and notifications (3c), as shown in Fig. 6:

1. First of all, the cloud node can receive external data from the domains of interest from multiple sources, (1) which are received through a message topic integrated within the ESB.
2. The cloud node will also obtain information from the fog nodes with relevant data to be processed. Such information is also received through a message topic (2) integrated within the ESB.
3. Once the information is transformed and processed by the CEP engine, detected situations of interest can be submitted to interested parties, such as the fog nodes which will be subscribed to a message topic (3a) or other parties which can receive, as previously mentioned, notifications (3b), services (3c) and also subscribe to a message topic themselves (3d).

## 5. Case study

IoT applications for healthcare are taking considerable relevance nowadays [48]; for this reason, we have selected this scenario as a case study for our proposal.

### 5.1. Description

Suppose we find ourselves in the scenario of a pandemic of a severe respiratory disease. Let E be the system's set of edge nodes $\{e_1,\ldots,e_n\}$, F be the system's set of fog nodes $\{f_1,\ldots,f_n\}$ and C the system's set of cloud nodes $\{c_1,\ldots,c_n\}$. Also, U will be the end user $\{u\}$. Given the set of nodes $N = E \cup F \cup C \cup U$, the message sent from the origin node $n_o$ to the target node $n_t$ will be represented as $n_o n_t$.

In this scenario, we could have a hospital equipped with the following nodes, as previously represented in Fig. 3:

- A series of edge nodes as described in Section 3.2: access devices ($e_1$), ventilators ($e_2$), windows ($e_3$), exterior emergency light ($e_4$), interior emergency light ($e_5$) and control panel with patient local history ($e_6$).
- The hospital would also have a fog node to control all the edge ones ($f_1$) and another one for patient admission and staff distribution ($f_2$).
- We will also have a cloud node ($c_1$) in which we will have pharmaceutical laboratories, pharmacies, Ministry of Health and other external data providers.
- Finally, the end user in communication with the edge through the fog will be the surveillance unit ($u$).

Fig. 7 represents the communications between edge nodes, fog ones

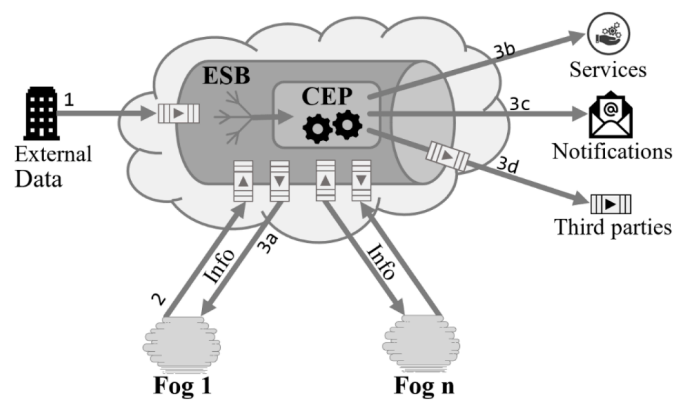

**Fig. 6.** Insight on Cloud Nodes and communications.

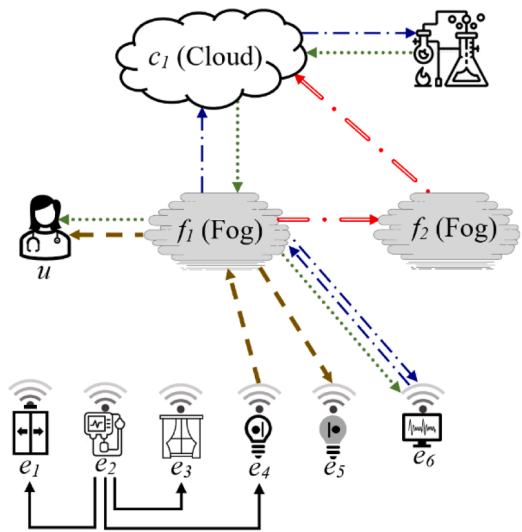

**Fig. 7.** Communications between Nodes in the given hospital scenario.

and cloud in this particular scenario.

Edge nodes in the same room would communicate their events to each other and have a set of rules programmed into them (see black lines in Fig. 7). For example:

- If the $O_2$ blood level detected by the ventilator is below a certain threshold, the agent in the external emergency light would detect it and immediately activate the light ($e_2e_4$), the door would be unlocked ($e_2e_1$) and the room would be ventilated ($e_2e_3$).

The fog node which controls edge devices in the rooms would receive the relevant information from all the edge nodes in each room on every floor (see brown lines).

- The fog node will receive a message any time a room activates the external light ($e_4f_1$). When the number of rooms on the same floor with an event that has activated the external light is above a certain threshold, it sends an alert to the hospital surveillance unit to immediately send more doctors to that floor ($f_1u$), at the same time it turns on the internal lights of the remaining rooms for a short period of time so the doctors inside them are aware of this fact ($f_1e_5$).

We could also have the patient history in the fog node: $e_6$ would send all the information recorded about the medicines and reactions to the fog node ($e_6f_1$) through the input topic, to be processed (see blue lines in Fig. 7):

- Once data is analyzed in the CEP engine, we could detect if the patient is being given some medicine that previously gave him/her a reaction or it can be seen if, for example, the same medicine is causing other patients any side effects. This is relevant data, not only in terms of changing the treatment, and therefore it can be sent to the control panel in the room ($f_1e_6$), but also in terms of research so it should be raised to the cloud ($f_1c_1$). An example of such a situation would be the one in Section 3.2, in which pharmaceutical laboratories could receive real-time information on the adverse effects of administering a drug (or vaccine, as in the example in Section 3.2).

On the other hand, the cloud can collect information from many sources, such as pharmaceutical laboratories, pharmacies, the Ministry of Health, and so on (see green lines in Fig. 7).

- For instance, it could be detected from the analysis in the SOA 2.0 of the information provided by pharmaceutical laboratories and pharmacies that the demand for a particular medicine administered by the fog hospital $f_1$ is growing out of control, which could lead to a lack of stock in a short period of time. That information should reach the hospital fog node ($c_1f_1$) and be transmitted to the surveillance unit in the hospital ($f_1u$) and should also be shown in the respiratory patients' control panels ($f_1e_6$). Besides, continuing with the example of adverse effects, pharmaceutical laboratories can alert in real time all hospitals where a particular drug is being administered that it is causing serious adverse reactions to certain patients.

The example could be extended to many other cases. For instance (see red lines in Fig. 7), the fog node in admissions and staff distribution ($f_2$) registers all incoming emergency patients with respiratory problems requiring specialized medical attention. This node will also be subscribed to $f_1$ to detect active red lights ($f_1f_2$) and could, in turn, call pulmonologists on duty at their homes depending on the number of new emergency patients and the amount of red lights per floor.

On the other hand, we could program several rules in this fog node ($f_2$) to send relevant information to the cloud nodes ($f_2c_1$). For example, notifications could be sent to the cloud with the number of pulmonologists in the hospital and the number of patients with severe respiratory symptoms. If the cloud receives this information from hospitals throughout the region, decisions can be made regarding staff redeployment, new contracts and health alerts for the population to observe symptoms or to advise them to go to the least collapsed centers, if necessary.

*5.2. Agent Rules and CEP Patterns*

In this section, we give some small code snippets to illustrate the kind of conditions we can use at each level of the architecture in order to obtain added knowledge through the use of CEP and agent-oriented software.

**Rules and situational-aware data at the edge**

Since several edge nodes are interested in $O_2$ saturation, such a value could be broadcast by the ventilator to all the edge nodes in the room, as shown in Listing 1.

**Listing 1.** Code snippet to broadcast measured $O_2$.

```
sensing.setSensorValue(value);
ACLMessage msg = new ACLMessage(ACLMessage.INFORM);
msg.setContent(value + "");
for (AID aid: agents) {
    msg.addReceiver(aid);
    }
O2SensorAgent.this.send(msg);
```

When the external light agent receives the $O_2$ saturation level and it is under 90%, it turns on the external light and sends a message to the fog topic, as shown in Listing 2. The fog will be in charge of counting the number of external lights on at a given time.

**Listing 2.** Code snippet for actions when $O_2$ is under 90%.

```
ACLMessage msg = myAgent.receive(mt);
if (msg != null) {
    try {
        setSensorValue(Integer.parseInt(msg.getContent()));
        agent.logMessage("O2 value " + msg.getSender().getLocalName()
            + ": " + msg.getContent());
        if (sensorValue <= 90) {
            myAgent.activateExternalLight();
            myAgent.notifyFog ()
            }
    } catch (NumberFormatException ex) {{…}}
}
```

**Patterns and context-aware actions programmed in the Fog**

When the fog receives external light on messages from the same floor at a certain period of time (see CEP pattern in Listing 3), and the number of external lights on exceeds a threshold —4, for example— (see CEP pattern in Listing 4), it sends an alert to the surveillance unit to immediately send more doctors to that floor.

**Listing 3.** ExternalLightByFloor pattern EPL code.

```
@Name("ExternalLightByFloor")
@Tag(name="domainName", value="Fog")
insert into ExternalLightByFloor
select current_timestamp() as timestamp,
   a1.floor as floor,
   count(a1.isOn) as count
from pattern [(every a1 = ExternalLight(a1.isOn = true))].win:time_batch(10
   minutes)
group by a1.floor
```

**Listing 4.** SurvaillanceUnit pattern EPL code.

```
Listing 4
@Name("SurveillanceUnit")
@Tag(name="domainName", value="Fog")
insert into SurveillanceUnit
select a1.timestamp as timestamp,
   a1.floor as floor
from pattern [(every a1 = ExternalLightByFloor(a1.count >= 4))]
```

**Data and context-aware patterns in the cloud**

It could be detected by the information provided by pharmaceutical laboratories and pharmacies that demand for a particular medicine that is administered by the fog hospital $f_1$ is growing out of control, which could lead to a lack of stock in a short period of time. To detect it, we receive information from several sources: pharmaceutical laboratories, pharmacies and hospitals. With pattern in Listings 5 and 6, we will detect when a particular medicine has very high demand (let's say more than 1000 units per hour). With the patterns in Listings 7 and 8, we detect a stock shortage of a particular medicine in pharmacies (let's say the stock is under 5 units per hour). With Listings 9 and 10, we detect the use of a particular medicine for respiratory diseases in a hospital. Finally, the pattern in Listing 11 detects when the medicine being used in the hospital for respiratory diseases has very high demand in pharmaceutical laboratories and stock shortage in pharmacies, therefore detecting a high risk of running out of stock for the medicine in question. That information should reach the hospital fog node and be transmitted to all doctors in that hospital and it should also be shown in the respiratory patients' control panels.

**Listing 5.** DemandByLaboratory pattern EPL code.

```
@Name("DemandByLaboratory")
@Tag(name="domainName", value="Fog")
insert into DemandByLaboratory
select current_timestamp() as timestamp,
   a1.id as id,
   a1.type as type,
   a1.place as place,
   count(a1.place) as count
from pattern [(every a1 = Medicine(a1.place = 'laboratory'))].win:time_batch(1
   hours)
group by a1.id, a1.type, a1.place
```

**Listing 6.** VeryHighDemandByLaboratory pattern EPL code.

```
@Name("VeryHighDemandByLaboratory")
@Tag(name="domainName", value="Fog")
insert into VeryHighDemandByLaboratory
select a1.*
from pattern [(every a1 = DemandByLaboratory(a1.count > 1000))]
```

**Listing 7.** StockByPharmacy pattern EPL code.

```
@Name("StockByPharmacy")
@Tag(name="domainName", value="Fog")
insert into StockByPharmacy
```

(*continued*)

```
select current_timestamp() as timestamp,
   a1.id as id,
   a1.type as type,
   a 1.place as place,
   count(a1.place) as count
from pattern [(every a1 = Medicine(a1.place = 'pharmacy'))].win:time_batch(1
   hours)
group by a1.id, a1.type, a1.place
```

**Listing 8.** StockShortageByPharmacy pattern EPL code.

```
@Name("StockShortageByPharmacy")
@Tag(name="domainName", value="Fog")
insert into StockShortageByPharmacy
select a1.*
from pattern [(every a1 = StockByPharmacy(a1.count <= 5))]
```

**Listing 9.** UseByHospital pattern EPL code.

```
@Name("UseByHospital")
@Tag(name="domainName", value="Fog")
insert into UseByHospital
select current_timestamp() as timestamp,
   a1.id as id,
   a1.type as type,
   a1.place as place,
   count(a1.place) as count
from pattern [(every a1 = Medicine((a1.place = 'hospital' and a1.type =
   'respiratory')))].win:time_batch(1 hours)
group by a1.id, a1.type, a1.place
```

**Listing 10.** RespiratoryUseByHospital pattern EPL code.

```
@Name("RespiratoryUseByHospital")
@Tag(name="domainName", value="Fog")
insert into RespiratoryUseByHospital
select a1.*
from pattern [(every a1 = UseByHospital(a1.count >= 1))]
```

**Listing 11.** MedicineStockBreak pattern EPL code.

```
@Name("MedicineStockBreak")
@Tag(name="domainName", value="Fog")
insert into MedicineStockBreak
select current_timestamp() as timestamp,
   a3.id as id
from pattern [(every (a1 = VeryHighDemandByLaboratory and a2 =
   StockShortageByPharmacy(a2.id = a1.id) and a3 = RespiratoryUseByHospital(a3.
   id = a2.id)))].win:time_batch(24 hours)
```

Please note that the pattern code can be defined manually or through the use of the MEdit4CEP [49] graphical tool, which permits automatic code generation. The graphical definition of the patterns in Listing 3 to 11 can be found at the Mendeley repository referenced in the supplementary material section.

## 6. Evaluation

As previously mentioned, in the past we evaluated an architecture integrating an ESB together with a CEP engine in a cloud node and results showed that the architecture had very good performance (see [6]).

In our present work, it is necessary to evaluate the performance of both fog and edge nodes. We have considered only one fog and one edge node to simplify the tests whilst looking for the limit of their capacity. For this purpose, we have employed (i) an Orange Pi Zero with an ARM Cortex-A7 Quad-Core (Allwinner H2) CPU, 512MB RAM, 8GB MicroSD Class 10 and 10/100 RJ45 Ethernet as the edge node and (ii) a Raspberry Pi 3 with Quad Core 1.2GHz Broadcom BCM2837 64bit CPU, 1GB RAM, a 32GB MicroSD Class 10 and 100 Base Ethernet as the fog node.

Particularly, we have conducted performance tests with the technologies and communications as proposed in our architecture for the fog and edge nodes (see Section 6.1). Then, the CEP and software agents technologies' performance have been evaluated separately, also alleviating the communications load on the architecture (see Section 6.2).

*6.1. Atmosphere Architecture Evaluation*

In order to obtain both nodes' performance rates and to stress the systems in order to check their limit with regards to the incoming event message rate that can be processed in real time, we have conducted the following procedure (see Fig. 8):

- For every test set, we submit the incoming event message rate from the edge node to the fog one through a message MQTT topic in the Mosquitto broker. Every message contains JSON information with random values for *IsOn* and *floor* (see case study in Section 5).
- Such messages are processed in the Esper CEP engine executed on the fog node and the resulting complex event message is then submitted back from the fog node to the edge one, through a message MQTT topic in a Mosquitto broker. Note that the edge node is in charge of processing it.
- Be aware that, in order to force the system, every incoming message in the fog node results in a complex event message to be sent to the edge node.
- In parallel, the edge node is sending a simulated humidity information message through the JADE platform every second.

We have developed and tested the system both with MQTT Quality of Service (QoS) 1 and QoS 0, and measured CPU usage and response time in both nodes. In this testing scenario, the response time refers to the time elapsed since the edge node sends a message by MQTT to the fog node, the latter processes it with the CEP engine and the output message from the CEP engine arrives back to the edge node. Therefore, note that every message implies in fact two communications between the nodes —the message going through MQTT from the edge to the fog and the one going through MQTT from the fog to the edge— and, always in parallel, the one going through the JADE platform from the edge to the fog gateway every second. Besides, when using QoS 1 (see Fig. 8, left-hand side), each message sent by MQTT implies a recognition message sent back to the sender. Therefore, when the edge node sends the initial incoming event message to the fog one, the latter submits one response back, and when the fog node sends the complex event message to the edge, the latter also submits one response in return, therefore, two more messages for every incoming message. When using QoS 0, no acknowledgment message is submitted (see Fig. 8, right-hand side).

The charts explained in the following lines show the results for incoming rates of 10, 20 and 30 events per second with QoS 1 and 200, 300 and 320 events per second for QoS 0; the system is not able to deal with higher rates of events. With the aim of finding the incoming rate limit, we initially did short tests (130 seconds) to see the evolution of CPU usage and response time.

As we can see in Table 1, when using QoS 1, CPU Usage remains stable on less than 20% of usage for the edge node —see Fig. 9— and 50% of usage for the fog node —see Fig. 10—, discarding the warm up. However, when we look closely at the response time in Fig. 11, it remains stable for incoming rates of 10 and 20 events per second and starts to fluctuate from 30 events per second. We can see that, at the rate of 30 events per second, response time fluctuates between approximately 20 and 45 milliseconds. This is due to the fact that, having a higher message rate, the CPU is less likely to decide to change the running process and therefore a context switching is not required. For this reason, in certain moments we can obtain better response times at the rate of 30 events per second than at 20, in the last scenario we understand that the processor is constantly making context switches with the resulting time cost. Despite this, response time remains below 50 milliseconds in all cases and the averages, as shown in Table 1, are under 35 milliseconds for the higher incoming rate.

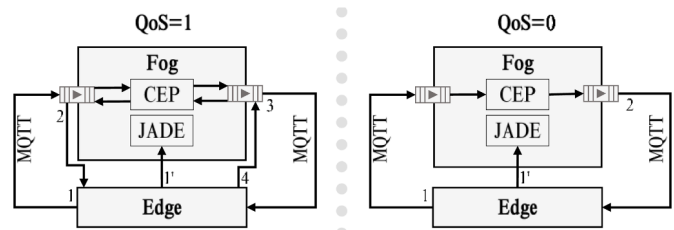

**Fig. 8.** Message flow between nodes during tests.

In addition, it is important to highlight that, as previously explained, each message actually corresponds to 3 messages in the system; therefore, these rates of 10, 20 and 30 messages per second could be considered as 30, 60 and 90 messages per second, respectively. Besides, when using QoS 1 — as is the case—, each message sent by MQTT implies a recognition message sent back to the sender, therefore 2 more messages for every incoming message. This would imply that the initial rates of 10, 20 and 30 incoming event messages per second are actually 50, 100 and 150 messages per second, respectively. We consider these rates and response times appropriate in the scope of IoT for QoS 1.

In any case, in multiple IoT domains we may not need QoS 1; therefore, we have repeated the tests with QoS 0 for incoming rates of 200, 300 and 320 events per second. Table 2 shows that, when using QoS 0, CPU usage remains stable at around 20% of usage for the edge node —see Fig. 12— and 40 to 70% of usage for the fog node —see Fig. 13—, discarding the system's warm up. When we analyze the response time (see Fig. 14), we can check that, although we have some peaks with higher response times, the average response time remains below 7 milliseconds; these are also shown in Table 2.

Additionally, it is important to remember that each message actually corresponds to 3 messages in the system given that it refers to the time elapsed since the edge node sends a message by MQTT to the fog node, the latter processes it with the CEP engine and the output of the CEP engine arrives back to the edge node and also, in parallel, a message is sent from the edge node to the fog one through JADE infrastructure. Therefore, these rates of 200, 300 and 320 messages per second could be considered 600, 900 and 960 messages per second. Since we are using MQTT with QoS 0, no acknowledgement message is sent back to the message sender in this scenario.

Once such a performance limit is found, we proceeded to test the system with the higher rate for a longer time and therefore with more incoming messages. In particular, we have performed tests with one million events. As detailed in Table 3, when using QoS 1 with an incoming event rate of 30 events per second, CPU usage remains stable at less than 20% of usage for the edge node —see Fig. 15— and less than 60% of usage for the fog node —see Fig. 16—, discarding the warm up. Looking closely at the response time in Fig. 17, we can see that it fluctuates but, as detailed in Table 3, more than 97% of the response time data are under 50 ms.

When we proceeded to test the system with QoS 0 with 1 million incoming events, we detected that the system could not deal with the rate of 320 incoming event messages per second, but did was able to cope with 300 incoming event messages per second. As we can see in Table 4, when using QoS 0 with an incoming event rate of 300 events per second, CPU usage remains stable on less than 20% of usage for the edge node —see Fig. 18— and around 60% of usage for the fog node—see Fig. 19—, discarding the warm up. Besides, in Fig. 20 we can check that the response time fluctuates but, as detailed in Table 4, more than 90% of the response time data are under 5 milliseconds.

Note that the spreadsheet data represented in the charts from Figs. 9 to 20 are available at the Mendeley repository referenced in the supplementary material section.

Table 1
Average CPU usage and response time for QoS 1.

| Incoming message rate (events/s) | Real message rate (messages/s) | Edge node average CPU usage (%) | | Fog node average CPU usage (%) | | Average response time (ms) |
| --- | --- | --- | --- | --- | --- | --- |
| | | Including warm up | Excluding warm up | Including warm up | Excluding warm up | |
| 10 | 50 | 15.5 | 11.6 | 42.7 | 43.6 | 8.5 |
| 20 | 100 | 16.3 | 12.6 | 42.3 | 43.5 | 45.9 |
| 30 | 150 | 18.3 | 12.9 | 44.5 | 46.3 | 34.2 |

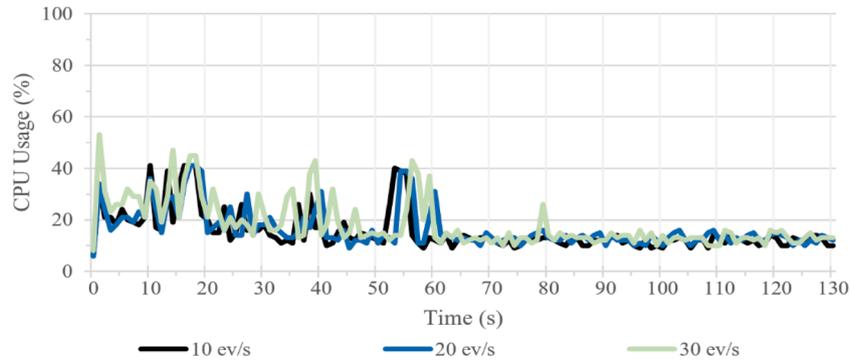

Fig. 9. CPU usage in the Edge Node with MQTT QoS 1.

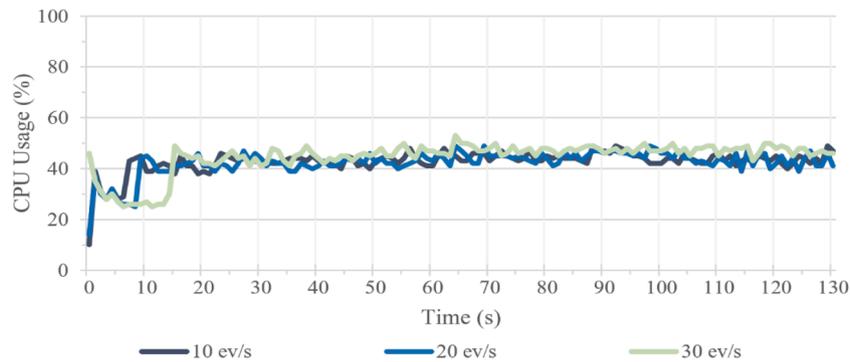

Fig. 10. CPU usage in the Fog Node with MQTT QoS 1

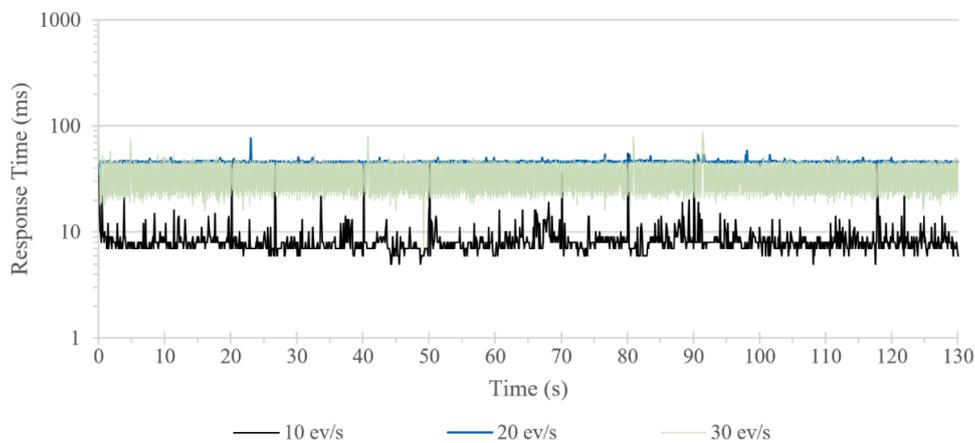

Fig. 11. Response time of communications between the Edge and Fog Nodes with MQTT QoS 1.

## 6.2. Evaluating CEP and software agents separately in the atmosphere architecture

In order to evaluate the suitability of the chosen technologies independently of the proposed architecture, we have carried out some additional performance tests of the main technologies in isolation, for the use of CEP in the fog node and for the use of software agents in the fog and edge nodes. This way, we can evaluate a straightforward deployment without communication between layers on the one hand, and a deployment where the edge nodes receive all data without being filtered by the CEP engine on the other, as explained in the following paragraphs. In these tests, we have focused on the QoS 0 evaluation as most of the MQTT communications between nodes are eliminated. In addition, we have selected the 300 incoming events per second rate tested in Section 6.1 for QoS 0.

**Table 2**
Average CPU usage and response time for QoS 0.

| Incoming message rate (events/s) | Real message rate (messages/s) | Edge node average CPU usage (%) | | Fog node average CPU usage (%) | | Average response time (ms) |
|---|---|---|---|---|---|---|
| | | Including warm up | Excluding warm up | Including warm up | Excluding warm up | |
| 200 | 600 | 27.9 | 15.6 | 44.3 | 45.7 | 6.9 |
| 300 | 900 | 26.8 | 15.4 | 52.3 | 57 | 6.3 |
| 320 | 960 | 30.9 | 18 | 46.5 | 51.3 | 5.2 |

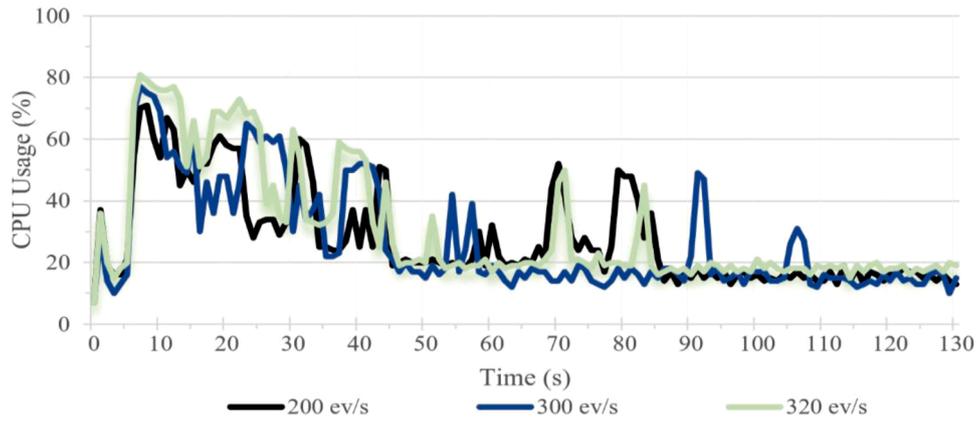

**Fig. 12.** CPU Usage in the Edge Node with MQTT QoS 0.

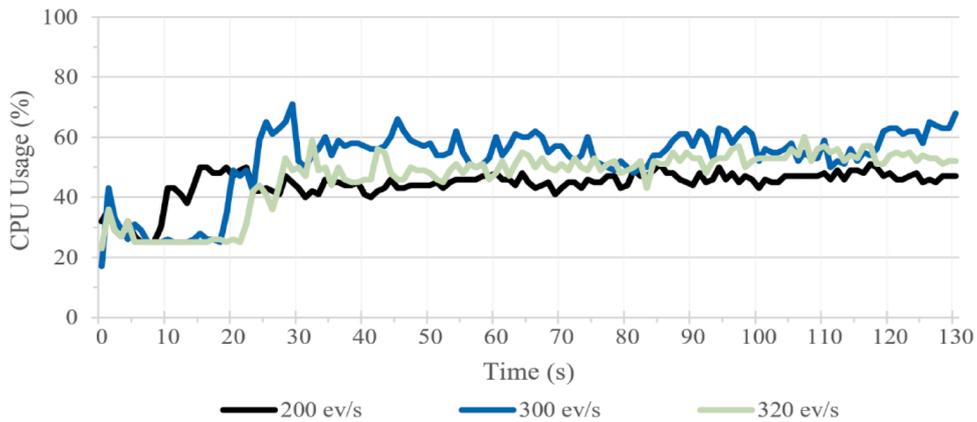

**Fig. 13.** CPU Usage in the Fog Node with MQTT QoS 0.

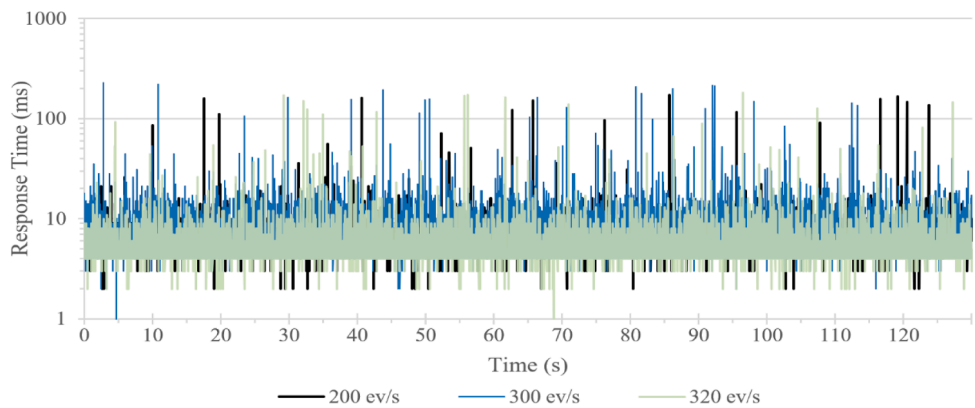

**Fig. 14.** Response time of communications between the Edge and Fog Nodes with MQTT QoS 0.

**Table 3**
CPU usage and response time distribution for QoS 1 for 1 million incoming events.

| CPU Usage Interval (%) | <=5 | 5-20 | 20-40 | 40-60 | >60 |
|---|---|---|---|---|---|
| Edge Node Sampling Percentage (%) | 0.26 | 99.48 | 0.22 | 0.04 | 0 |
| Fog Node Sampling Percentage (%) | 0 | 0 | 0.1 | 99.88 | 0.01 |

| Response Time Interval (ms) | <=5 | 6-10 | 11-50 | 51-100 | >100 |
|---|---|---|---|---|---|
| Sampling Percentage (%) | 0 | 0.21 | 97.81 | 1.58 | 0.4 |

Firstly, performance tests were carried out keeping only the CEP engine processing in the fog node; this has been called the *CEP-only* test. That is to say, we have not run JADE nor has there been any communication between the node in the fog and the one in the edge, apart from the fact that the event simulator has been kept on the edge node, as shown in the left-hand side of Fig. 21. The simulated data was sent, as previously said, through MQTT QoS 0 with a ratio of 300 events per second and all these data were processed in the CEP engine. The results

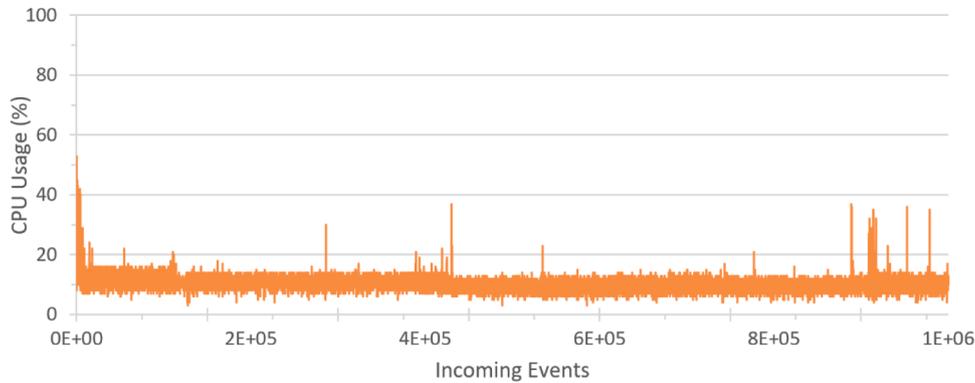

**Fig. 15.** CPU Usage in the Edge Node with MQTT QoS 1 and one million incoming events.

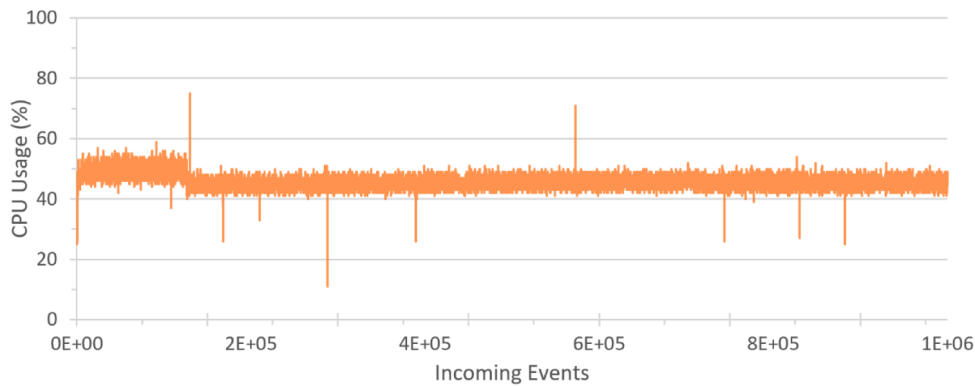

**Fig. 16.** CPU Usage in the Fog Node with MQTT QoS 1 and one million incoming events.

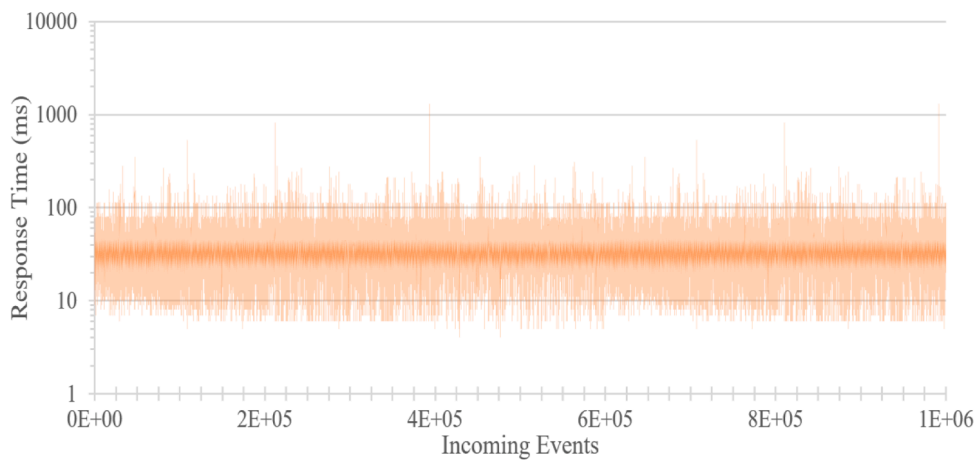

**Fig. 17.** Response time of communications between the Edge and Fog Nodes with MQTT QoS 1 and one million incoming events.

Table 4
CPU usage and response time distribution for QoS 0 for one million incoming events.

| CPU Usage Interval (%) | <=5 | 5-20 | 20-40 | 40-60 | 60-80 | >80 |
|---|---|---|---|---|---|---|
| Edge Node Sampling Percentage (%) | 0.06 | 96.58 | 1.95 | 0.9 | 0.48 | 0.03 |
| Fog Node Sampling Percentage (%) | 0 | 0 | 0,84 | 80,25 | 18,91 | 0 |
| Response Time Interval (ms) | <=5 | 6-10 | 11-50 | 51-100 | >100 | |
| Sampling Percentage (%) | 90.57 | 6.4 | 2.56 | 0.23 | 0.25 | |

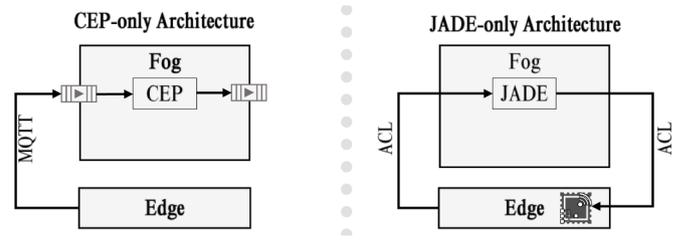

Fig. 21. Message flow between Nodes during *CEP-Only* (left-hand side) and *JADE-Only* (right-hand side) tests.

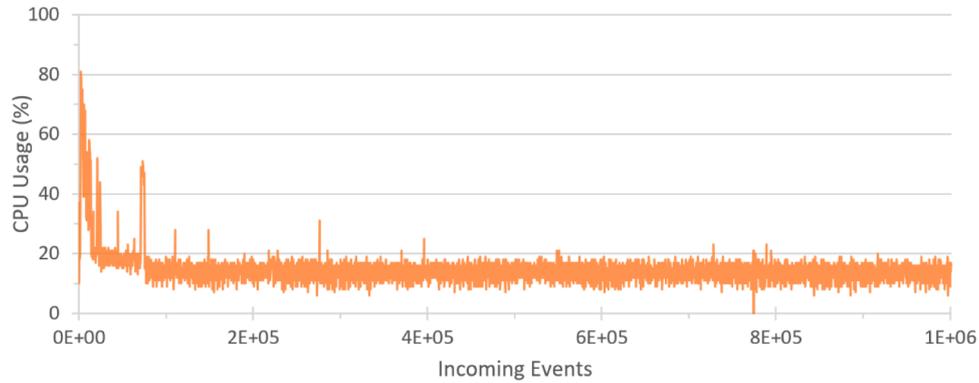

Fig. 18. CPU Usage in the Edge Node with MQTT QoS 0 and one million incoming events.

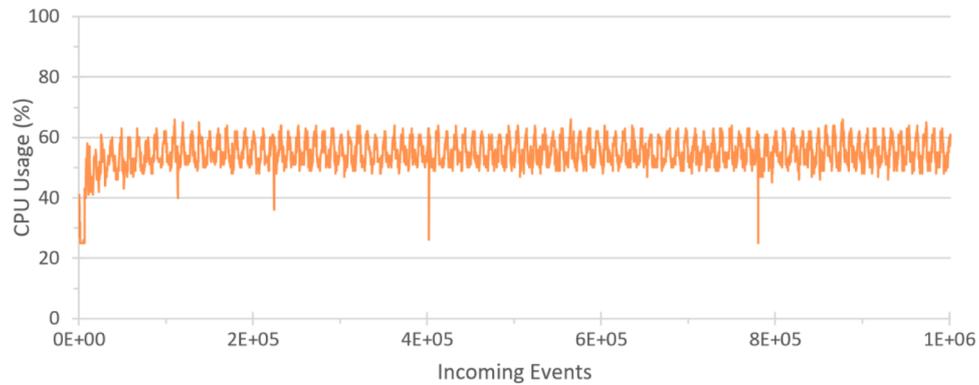

Fig. 19. CPU Usage in the Fog Node with MQTT QoS 0 and one million incoming events.

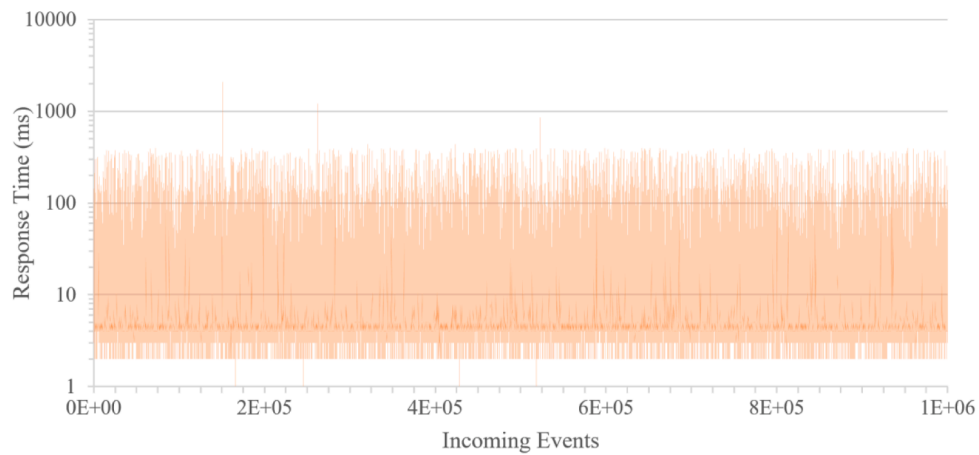

Fig. 20. Response Time of Communications between the Edge and Fog Nodes with MQTT QoS 0 and one million incoming events.

show that the fog node is very fast; as expected, there is faster event processing with lower CPU consumption than in the Atmosphere architecture since we have removed all communications between nodes except to send data from the simulator. In Fig. 22, we show the CPU usage and response time values for the fog node in our original test case, compared with the values obtained with this new *CEP-only* configuration. With this new configuration, the CPU usage remains below 30% and response times around 7 ms, as we can see in the left and right-hand side of Fig. 22, respectively. This shows how powerful the CEP engine can be in a fog node and how the performance of the Atmosphere architecture is not particularly affected when combining the use of CEP with the use of software agents, despite the increase in the number of communications between nodes.

Secondly, we have performed tests removing the CEP engine and keeping all the processing in the JADE; this has been called the *JADE-only* test. In this case, we still keep the data simulator in the edge node, the JADE gateway in the fog node and a software agent in the edge node. All data sent by the simulator, this time through ACL and not through MQTT, are processed by a service at the JADE gateway and, once processed, sent to the agent at the edge node, as shown in the right-hand side of Fig. 21. As we can see from the test results, the CPU consumption in the fog node (left-hand side of Fig. 23) when only using JADE is also very good —under 40%—, even less consumption than our original architecture, while it is higher on the edge node (right-hand side of Fig. 23) since it remains around 35% versus the original 10%: as expected, the edge device is overloaded in these tests when a large number of messages is being submitted to it. Response times (see Fig. 24) for the *JADE-only* configuration are also lower than in the proposed Atmosphere architecture, but we have to bear in mind that we are reducing the number of communications when measuring the response time. Note that the submission of humidity has been removed from these tests as well as the submission through MQTT from the edge node to Mosquitto and from Mosquitto to CEP and vice versa, which have been replaced by a unique message from the edge node to the JADE gateway in the fog and the message back from the latter to the software agent in the edge node.

As a conclusion, we have checked that the JADE gateway in the fog node can handle a high number of events, but when distributing them, without prior filtering of unnecessary events, it can overload the node at the edge. Therefore, by combining the use of CEP in the fog node for filtering events coming from different nodes with the use of software agents for decision making of relevant data at the edge node, we will have an efficient and scalable system that facilitates communications between the different layers.

## 7. Related work

Although we can find multiple works that deal with edge computing, fog computing, edge-fog, fog-edge and so on, the vast majority of them approach it from a network communications and communication protocols perspective. We have found few works that study the software architecture that allows developers to take better advantage of the data flowing between such devices and the result of analyzing them. In the following lines, we describe the works we have found in which we appreciate a greater similarity to our proposal in terms of the problem to be solved, yet we will see that the approaches are rather diverse.

Zhao et al. [2] propose an IoT edge computing-enabled collaborative tracking architecture with three layers: the lower layer is the one composed by sensors, actuators, microprocessors and so on; the middle layer (edge) is conformed by ARM-based Raspberry Pi computational boards and smart devices and communicates with the former by Bluetooth and Zigbee. Finally, in the top layer we can find the cloud services, which communicate with the middle layer by Narrow Band (NB)-IoT and 4G. The edge layer is in charge of collecting, filtering, analyzing, storing and transmitting the data from one layer to the other. The main difference with our proposal is that no computing is performed on the layer where the sensors and actuators are, but all the information collected from the lower level is processed in their edge computing device. Besides, they do not perform any rules to detect situations of interest, but simply track the manufacturing materials.

Farahani et al. [4] propose a three-layer architecture to deliver real-time health monitoring and anomaly detection. The lower layer in the architecture is the one composed of the endpoint IoT health devices, the middle one is the one they call edge/fog computing layer, and the upper layer is the cloud one. In the device, they perform noise removal from an electrocardiogram (ECG) signal and extract the relevant features from them, then depending of whether the result of the prediction made by the algorithm applied to the device data exceeds the threshold, edge computing may or may not be required. Fog allows more computing tasks and more complex machine learning algorithms to take place on the edge nodes, thus supporting decision making with more complex models. Further support is also given by the cloud node, which will also be involved in heavier tasks such as model training or personalization. Even though the proposal deserves our attention and the proposed three-tier architecture bears some resemblance to the one we present in this paper, they only focus on machine learning models and their accuracy at the three different levels. However, we propose an architecture in which situations of interest can be detected in real time at any of the three levels and which may be of interest to the others, so that we enable two-way communication throughout the components of the architecture.

Also, the proposal from Bade et al. [50] deserves special mention due to their usage of both agent-oriented software and CEP. They propose a three-layer application middleware for post-processing of Wireless Sensor Networks (WSN) and RFID. Their proposed three layers—network edge, network layer and application layer—, would be equivalent to the edge, the fog and the final user application in our proposal. They propose the use of agents to filter, aggregate, transform and enrich data in the network edge. Such enriched data is then submitted to the network layer where it is processed by the CEP engine and relevant complex events detected are then submitted to the application layer, in so doing, informing the final user. Bidirectional communication is supported since applications can send processing instructions, queries and configurations to the network edge, where agents also support the execution of the business workflows of the domain in question.

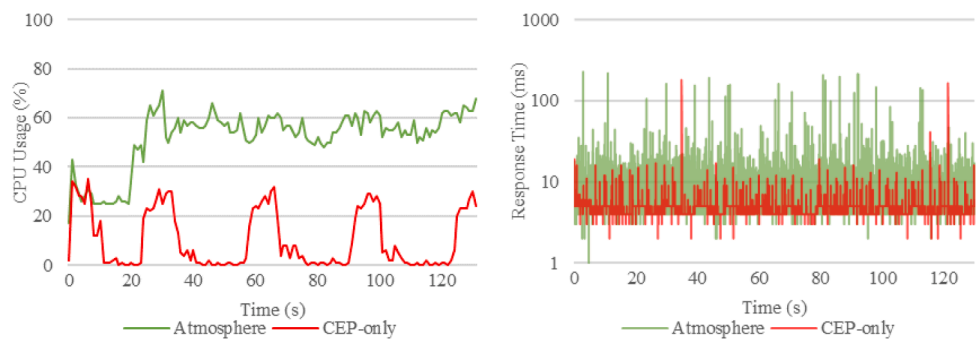

**Fig. 22.** CPU usage and response time in the Fog Node with MQTT QoS and 300 incoming events/s for atmosphere and *CEP-only* architectures.

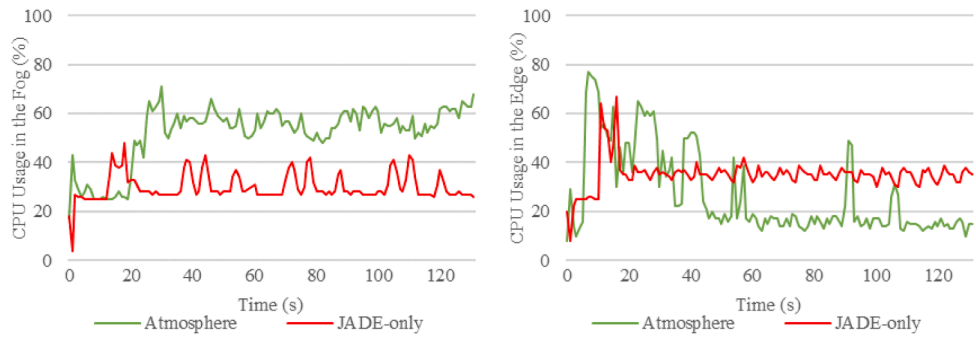

**Fig. 23.** CPU Usage in the Fog and Edge Nodes with MQTT QoS and 300 incoming events/s for atmosphere and *JADE-only* architectures.

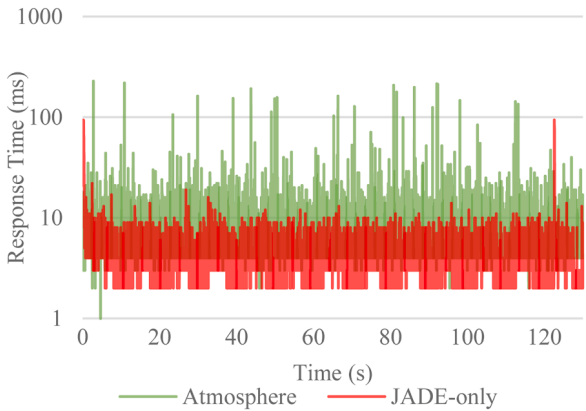

**Fig. 24.** Response time in the Fog Node with MQTT QoS and 300 incoming events/s for atmosphere and *JADE-only* architectures.

Also interesting is the proposal from Ai et al. [3]. Even though they mainly focus on network communications, it is worth noting how they visualize architectures where several edge nodes can exchange information with each other. In addition, each edge node can communicate with a fog node, which in turn communicates with more fog nodes horizontally and vertically, the fog node at the top of the communication pyramid being the one that connects with the cloud.

Shah-Mansouri and Wong [51] study the computing resource allocation in a fog-cloud computing paradigm with the aim of minimizing the power consumption of fog nodes and cloud servers of different IoT applications as well as determining part of the computation workload of each fog node that should be offloaded to the remote cloud servers for IoT users to obtain a higher quality experience.

The work from Debauche et al. [35] presents a three-tier architecture. The sensors or wearable devices send the information to the smart gateway, where they are anonymized, semantically enriched, compressed and encrypted before being submitted to the cloud where the analytical processing is done. Therefore, in this work no events of interest are detected at the edge or at the fog, but all the processing for the detection of situations of interest is done in the cloud causing the system to be slower and less scalable.

In [36], Maleki et al. present a 3-layer architecture, where the IoT devices send their data to the fog. Data are then processed in the fog in real time using Spark. In addition, the system has a cluster of fog nodes and a master that allows the streaming distributed processing. Only relevant events are sent to the cloud for storage.

The work presented by Hossain et al. [52] consists of a proposal in which raw data is sensed and submitted to the close edge device. This edge device supports the processing of such data and infers a situation, understanding by situation a set of information about an environment.

Some of the examples given are high temperature or low humidity. Then, such situations of interest are sent to the cloud where they can be further processed and where customers can subscribe to possible notifications.

Mahmud et al. [37] present a three-layer architecture to integrate sensors, fog computing and cloud computing. They have, on the one hand, health sensors that send the obtained information to the fog infrastructure, where a series of microcomputer instances processes received data, providing it to the services in the fog, which in turn feeds the health applications that would in turn allow the sensors' demand to be replenished. Communication between the fog and the cloud would also be bidirectional; in the cloud, the data would be processed in the virtual machine by the cloud's own services, to provide them to the applications in the cloud itself. However, they do not mention what type of technologies would be used for processing in both the cloud and the fog.

In Table 5, we summarize a set of features of the proposals, all of which are relevant in the scope of C-IoT. If these implement the feature in question, the + symbol is used, if they do not, we have used–, and / is used if it is partially provided. The examined features are the following:

A. 3-tier: the proposed software architecture is based on three tiers, namely edge, fog and cloud.

B. Edge device computing: a computing mechanism is provided at the edge tier.

C. Fog device computing: a computing mechanism is provided at the fog tier.

D. Cloud computing: a computing mechanism is provided at the cloud tier.

E. Detecting real-time situations: the computing mechanisms can provide real-time detection of situations in all the implemented layers.

F. Communicating real-time situations among tiers: the architecture permits communicating detected situations among the three tiers.

G. Communicating edge nodes with end users: a mechanism is provided to facilitate communication from the edge to the end user.

H. Multidomain: the proposed architecture is not limited to a particular domain.

I. Bidirectional communication: communication among the tiers is

**Table 5**
C-IoT feature implementation by analyzed proposals.

| Evaluated features | A | B | C | D | E | F | G | H | I | J |
|---|---|---|---|---|---|---|---|---|---|---|
| Zhao [2] | + | - | + | + | – | - | + | - | - | - |
| Farahani [4] | + | + | + | + | + | - | - | + | - | + |
| Bade et al. [50] | / | - | + | - | + | / | + | + | + | - |
| Ai [3] | + | + | + | + | / | / | - | + | - | - |
| Shah-Mansouri [51] | - | - | + | + | - | - | - | + | - | - |
| Debauche [35] | + | - | + | + | / | - | - | + | / | / |
| Maleki [36] | + | - | + | + | + | / | + | + | - | + |
| Hossain et al. [52] | - | + | - | + | + | + | / | + | - | - |
| Mahmud [37] | + | - | + | + | + | + | - | + | + | - |
| Our proposal | + | + | + | + | + | + | + | + | + | - |

bidirectional.

J. Machine learning: the prediction of situations of interest is facilitated by machine learning techniques.

There are other works that cannot be compared with ours because they focus on other relevant issues in this same domain, but they will generally be complementary approaches. For example, very interesting is the proposal from Masri et al. [53], where a communication algorithm between devices in the fog is presented to distribute the processing load between different nodes in a collaborative way. This is not our objective, but our architecture could benefit from using an algorithm for this purpose in a complementary way.

To sum up, we can say that, although all proposals provide some features or others, few allow for the processing, real-time and bidirectional communication in the three tiers, which is the great challenge addressed in this paper. Without a doubt, something that we must face in our future work is the integration with machine learning techniques. We believe that this integration is feasible since we have approached it in other works with SOA 2.0 [7,54], but in this case, we will have to consider to what extent it is applied to each level of the architecture.

## 8. Discussion

The proposed architecture has proven to enable C-IoT with three tiers—edge, fog and cloud-, with multiple stakeholders and a wide range of applicability domains going one step further in the state of the art, taking advantage of the three tiers.

On the one hand, with edge computing we will benefit from better network performance and less latency since IoT edge devices can intelligently process data locally thanks to software agents. Besides, they can transmit relevant information to nearby fog devices, when useful. This way, this information can be used for other relevant functions in this or other layers of the architecture.

On the other, the location of fog devices between the edge nodes and the cloud provides better performance, distributing processing and enabling real-time response without high latencies. Moreover, fog nodes enable the system to scale more easily, grouping the edge nodes according to the needs of the real-world case study. Besides, fog nodes can perform data aggregation to send partially processed data to the cloud, thus facilitating the addition of several sources in such cloud nodes. Thanks to these two features, the design and deployment of different configurations in terms of the number of nodes per level and their grouping will allow developers to deal with different scenarios in a more efficient way, adapting to the needs of the case study in question.

Finally, cloud nodes can process huge amounts of data and send already processed relevant data to the fog nodes, thus alleviating the computing load in them.

As a result, we have time-sensitive distributed processing which allows real-time decision making at each layer of the architecture according to local information and external pre-processed information specially filtered for the domain in question.

Our research questions, presented in Section 1, are answered below.

RQ 1. Can we integrate various specific software approaches that have proven to be efficient at the edge, fog and cloud, respectively, to provide a complete three-tier software architecture in which all three tiers can process information?

Yes, we have shown how we can integrate various specific software approaches that have proven to be efficient at the edge, fog and cloud, respectively. In particular, we have proposed the use of agent-oriented programming as an efficient solution for making intelligent decisions based on sensed data and data shared between the different edge nodes of the architecture [5], and the use of a SOA 2.0 with CEP for real-time heterogeneous data processing and notification in the cloud [8].

Besides, even though we already proved the efficiency of combining SOA 2.0 and CEP also for a fog node [8], we have chosen to propose a lighter and faster alternative, using CEP without the ESB, since, in the fog, we will only integrate messages coming from our platform and, therefore, they will already be in a homogenous and appropriate format. Moreover, not only the input but also the output will always be addressed to a MQTT topic, so we will not need to adapt the communications protocol either.

RQ 2. Would it be possible to have real-time data processing at every level of the integrated architecture as well as two-way communication between the edge and fog nodes and the fog and cloud ones by adjusting to system resources?

Yes, we have ensured that the three tiers process the information according to their capacities and resources: the edge is processing the information autonomously with the software agents, the fog is processing it with the CEP engine and the cloud is doing so with the SOA 2.0 combined with the CEP engine. We also ensure a collaborative architecture multi-dimensionally. On the one hand, we have the collaboration between all edge nodes that share information with each other; similarly, if we wanted to have several fog nodes in the system they could also share information with each other and the same would happen with the cloud nodes. On the other hand, we share information from the edge to the fog and vice versa, and from the fog to the cloud and vice versa; therefore the relevant information is shared by the three tiers. Furthermore, when having multiple actors in our case study, we are sharing information between users (patients in the edge), state institutions (hospitals in the fog), government agencies (Ministry of Health in the cloud), companies (pharmaceutical laboratories in the cloud), research institutions (research groups in the cloud) and so on and so forth. Furthermore, the performance tests carried out show how the devices with the least resources, in the fog and on the edge, are perfectly capable of processing data in real time for a rate of input data according to an IoT scenario.

RQ 3. If the answer to RQ2 is affirmative, would it be useful to have two-way communication among the three system layers and data processing in all of them in a broad range of IoT scenarios?

Yes, it would. We have shown an example that clearly illustrates the usefulness of the proposal in a hospital environment and we have also illustrated the C-IoT domain with the amenities and energy management and monitoring in a hotel. As these scenarios, many other IoT environments could benefit from this architecture that allows the processing and exchanging of situations of interest detected at the three levels of the architecture in order to have greater contextual and situational knowledge for decision making inside a community or even among countries [55]. Undoubtedly, the different actors in society (administrations, public and private institutions, citizens and industry) are increasingly prone to share data and encourage this collaboration, which provides benefits for everyone. With this architecture, we hope to move a step forward in this direction towards a society capable of taking better advantage of IoT in multiple domains and therefore conducting better decision making on a daily basis for greater economic benefit and social welfare.

RQ4. Besides, if the answer to RQ2 is affirmative, can we ensure interoperability between layers and coordination between devices of the same or different layers despite the lack of agreement on communication standards for the IoT, and can we do it in a secure way?

Yes, secure communications can be provided to ensure interoperability between layers and coordination between devices of the same or different layers. First of all, agents communicate through JADE platform; secondly, the communications between the different levels are done through a Mosquito broker using an MQTT protocol. Concerning JADE platform, it provides support for security in MAS systems; in particular, it provides features for authentication and permissions and message integrity and confidentiality. Regarding the security for communications between nodes, Mosquitto provides authentication options through certificates or pre shared keys with SSL/TLS as well as access control restrictions. Of course, we might use another message broker and/or communication protocol in a different implementation of the architecture, but it is expected that chosen brokers and protocols offer security support mechanisms. Finally, the ESB in the cloud also provides security features, which indeed is a particular feature of an ESB.

On interoperability, as previously mentioned, although there is currently no agreement on the standards to be used in terms of communications protocol and message payload format for the IoT, there is a trend towards the use of messaging brokers such as Mosquito and its MQTT protocol and JSON format for message payload format. We can add a transformation module, where necessary, to send the data to the corresponding node in the required format, as explained in Section 4.2.

Concerning coordination between devices of the same and different layers, on the one hand, we have previously highlighted that JADE provides a specific mechanism for agent coordination; on the other, message brokers in general, and Mosquitto in particular, provide different configurations for communications so that several coordination models are supported.

## 9. Conclusions

In this paper, we have taken another step forward in the state of the art of software architectures to facilitate C-IoT. The proposed technological solution makes use of three paradigms and specific technologies (agent-oriented software, SOA 2.0 and CEP) for the three layers of C-IoT (sensing, gateway and services), which will be carried out by processing real-time data at the edge, fog and cloud, respectively. The novel software architecture proposed and deployed along the three tiers not only adapts to the needs and processing capabilities of each tier, but also permits interoperability and two-way communications between the three levels and data processing in all of them, multiplying the knowledge in each tier and therefore significantly improving collaborative decisions in real time. Besides, being able to process data at the edge will translate into being able to make faster decisions at the edge when information from other layers is not relevant for the decision-making process, therefore being able to act more quickly when the scenario conditions require it. Not only this, but we also avoid overloading the networks and the computation in other layers with information that does not need to be processed in them. Furthermore, when processing data in the fog, it is not only saving the time required to send data to the cloud which could delay communications and decision-making and actions, but also the fact that, in certain scenarios, we may encounter privacy restrictions that can make it difficult to send such information to the cloud, requiring further processing to anonymize the data, and its consequent delay, before it is sent. Keeping and processing such data in the fog provides further security to the data collected in our architecture.

As part of our future work, we expect to analyze which machine learning mechanisms and algorithms would be the most suitable for the system and at which levels it would be reasonable to integrate them to provide a competitive advantage without damaging system performance. Machine learning techniques will make it possible not only to detect situations of interest but also to predict them based on the data processed in real time, and to anticipate decision-making based on these predictions, which can be key in certain application domains to prevent unwanted situations. Furthermore, in our future work we will study the benefits that the proposed architecture based on the integration of agent-oriented software and CEP can provide in the field of Industry 4.0. Our proposed system should integrate all elements that depend on the production chain by connecting all production machines and sensors through the IoT, connecting its enterprise resource planning system with the control system in manufacturing, operations, marketing, warehousing, distribution and delivery and permitting an improved low latency decision making.

## CRediT authorship contribution statement

**Guadalupe Ortiz:** Conceptualization, Methodology, Writing – original draft, Writing – review & editing, Funding acquisition. **Meftah Zouai:** Software, Investigation, Writing – review & editing. **Okba Kazar:** Conceptualization. **Alfonso Garcia-de-Prado:** Validation, Visualization. **Juan Boubeta-Puig:** Software, Writing – review & editing.

## Declaration of Competing Interest

None.

## Acknowledgements

This work was partially supported by the Spanish Ministry of Science and Innovation and the European Regional Development Fund (ERDF) under project FAME [RTI2018-093608-B-C33] and excellence network RCIS [RED2018-102654-T]. We also thank Carlos Llamas Jaén for his support with the setting up of the performance evaluation tests.

## Supplementary materials

We have uploaded our supplementary materials to Mendeley Data at the following dataset:

Ortiz, Guadalupe; Zouai, Meftah; Kazar, Obka; Garcia-de-Prado, Alfonso; Boubeta-Puig, Juan (2021), "Dataset for Atmosphere: Context and Situational-Aware Collaborative IoT Architecture for Edge-Fog-Cloud Computing", Mendeley Data, v1. URL: https://data.mendeley.com/datasets/mxtwc6vb8w/1.

The supplementary material is composed of the following folders and items:

Case Study Patterns: it includes the CEP patterns from Listing 3 to Listing 11, patterns in both EPL code and its graphical representation modeled with the MEdit4CEP tool.

Performance Evaluation Results Data Files: it includes the spreadsheets with CPU usage and response time values obtained from Atmosphere performance tests in Section 6.1.

**Guadalupe Ortiz** obtained her PhD in Computer Science at the University of Extremadura (Spain) in 2007, where she worked from 2001 as Assistant Professor. In 2009 she joined the University of Cádiz (UCA) as tenured Associate Professor in Computer Science and Engineering. Her research interests focus on the integration of complex-event processing and context-awareness in service-oriented architectures in the Internet of things.

**Meftah Zouai** obtained his PhD in Computer Science at the University of Biskra (Algeria) in 2021. In 2016, he obtained a Master in Networks Technologies and Communication. His research interests focus on Multi-agent and embedded systems and the integration of cloud computing and context-awareness in service-oriented architectures in the Internet of things.

**Okba Kazar** is full professor at the Computer Science department of the University of Biskra and director of LINFI laboratory. His main research fields are artificial intelligence, multi-agents systems, big data, Internet of Things and smart health. He has published more than 320 papers in international journals and conferences and two books: "Manual d'Intelligence artificielle" and "Big data security".

**Alfonso García-de-Prado** received his Ph.D. degree in Computer Science at the University of Cádiz (UCA), Spain, in 2017, where he has been an assistant professor since 2012. Previously he had been a developing programmer, analyst and consultant for international industry partners. His research focuses on context-aware service-oriented architectures, as well as their integration with complex event processing and the Internet of things.

**Juan Boubeta-Puig** is a tenured Associate Professor with the Department of Computer Science and Engineering at the University of Cádiz (UCA), Spain. He received his Ph.D. degree in Computer Science from UCA in 2014. His research interests include real-time big data analytics through complex event processing, event-driven service-oriented architectures, Internet of Things and model-driven development.